\documentclass[twocolumn,pre,floatfix,showpacs]{revtex4}
\usepackage{dcolumn}
\usepackage{graphicx}
\usepackage{graphicx,amssymb,amsmath}
\usepackage{natbib}
\usepackage{color}
\usepackage[normalem]{ulem}
\newcommand{\bra}[1]{\langle #1|}
\newcommand{\ket}[1]{|#1\rangle}

\begin{document}

%
%
%
%

\title{Intermolecular adhesion in conjugated polymers:  The role of the band gap and solitonic excitations}
\author{Jeremy D.~Schmit}
\affiliation{Department of Pharmaceutical Chemistry, University of
California,
 San Francisco, California 94158}\email[]{schmit@maxwell.ucsf.edu}

\author{Alex J.~Levine}
\affiliation{Department of Chemistry \& Biochemistry and
The California Nanosystems Institute\\
UCLA, Los Angeles, CA 90095-1596} \email[]{alevine@chem.ucla.edu}

\date{\today}

\begin{abstract}
Conjugated polymers are soft, one-dimensional conductors that admit
complex interactions between their polymeric, conformational degrees
of freedom and their electronic ones. The presence of extended
electronic states along their backbone allows for inter-chain
electronic tunneling at points where these polymers make near
passes. Using a combination of analytic modeling and Hartree-Fock
numerical calculations, we study the localized electronic states
that form due to such close encounters between semiconducting
conjugated polymers and explore how these states lead to
chain--chain binding. We also study the interaction of these
inter-chain binding sites with solitonic excitations on the chains.
From these results and a modified Poland-Scheraga model, we
determine the equilibrium structures of paired-chains formed by
intermolecular electronic tunneling. We calculate the energetic
ground state of such pairs and show the effective thermal
persistence length of the paired chains can vary over an order of
magnitude due to the intermolecular binding mechanism.
\end{abstract}

\maketitle

\section{Introduction}
\label{sec:intro}

Conjugated polymers~\cite{Su:79,Su:80,Takayama:80} can be thought of
as soft, low-dimensional conductors or semiconductors. Due to their
promise in technological applications such as LEDs, solar
cells~\cite{Heeger:98,solar,Yu2:95}, and even
biosensors~\cite{Chen:99,Gaylord:02}, a great variety of such
materials have been synthesized and studied. Independent of their
specific chemical details, the existence of extended electronic
states along their polymeric backbone provides for their surprising
(for a polymer) electronic properties. We refer to these
(semi-)conductors as ``soft'' since their polymeric backbones
typically have a short (nanometer scale) thermal persistence length;
the conduction path along the polymers in equilibrium solution
should be considered to be a tortuous random-walk.

These molecules admit strong correlations between their fluctuating
local configurational and electronic properties~\cite{Hone:91}. In
the extreme case, sharp (i.e. localized) bends in the polymeric
backbone result in regions of poor electron transport at these
so-called conjugation breaks.  Conversely, the electronic degrees of
freedom of the molecule influence its conformational degrees of
freedom. High electronic mobility along the thermally fluctuating
backbone's random path allows for complexities in the interactions
of a conjugated polymer with itself and with other such molecules.
For example, in conducting chains the pressure of the electron gas
trapped between nearby conjugation breaks (sharp bends) in the chain
should force such sharp bends apart and thus enhance the statistical
weight of locally straighter backbone contours~\cite{Pincus:87}.
Thus, the electronic degrees of freedom contribute to the effective
persistence length of the polymer. Moreover, looped conformations of
the polymer backbone generating close contacts of parts of the
polymer separated by a large arc length along its backbone may
further perturb the electronic structure of the molecule by
introducing tunneling between distant sites along the chain. This
sort of ``bridge conduction''~\cite{Hone:01} leads to localized
attractive interactions of the polymer at such crossing points where
the chain makes a near pass to either itself or another. Based on
this electronically mediated intra-molecular attraction, one may
imagine that conjugated polymers can self-aggregate into compact
globules in solution~\cite{Hone:01}. At higher polymeric
concentrations, one would expect to observe interchain binding, the
subject of this article.

More recently, detailed numerical calculations have been preformed
that determined the magnitude of these inter-chain tunneling matrix
elements for specific conjugated polymers~\cite{Bredas:02}. Using
these results and a simple tight-binding model for
metallic~\cite{Schmit:05} and/or semiconducting~\cite{Schmit:08}
chains, we showed that, for the case of pairs of chains,
intermolecular tunneling leads to attractive interactions for which
the binding energy is $\sim k_BT$. The tunneling creates a pair of
electronic bound states localized at the inter-chain crossing,
leading to a total decrease of the electronic contribution to the
system's energy on this scale. One may look at this attractive
interaction as an analogue to the traditional covalent chemical
bond, but at an energy scale two orders of magnitude weaker. We
understand the weakness of the bond to result from a combination of
molecular geometry creating a larger than typical internuclear
distance in the bond and the energetic cost of localizing the
electrons from the extended states of the molecule in order to fill
the new localized bound state. The magnitude of this intermolecular,
short-ranged interaction suggests that the statistical mechanics of
intermolecular binding driven by this mechanism is rather subtle,
since both chain configurational entropy and chain bending energies
contribute to the free energy of the system at the same energy
scale.

In the current article we study in more detail the effect of
inter-chain tunneling sites on the electronic states of
semiconducting chains by exploring analytically the interaction of
pairs of such tunneling sites, the interaction of solitonic chain
excitations with such tunneling sites, and determining the
electronic ground state of pairs of chains bound uniformly along
their length. We also use numerical calculations incorporating more
details of the chemical structure of the molecules in question to
study the dependence of the electronic binding interaction on the
local geometry of the crossing point between the two chains.
Finally, we use these results as input to a generalized
Poland-Scheraga model describing the binding of two chains. Using
this model, we determine the critical concentration for chain
pairing and study how the conformational statistics, as
parameterized by the effective thermal persistence length of the
paired chains evolves as a function of the strength of the binding
interaction.

To introduce the bandgap at the Fermi level and thus create the
semiconducting state, we use the SSH model~\cite{Su:79} for
polyacetylene. This simple tight-binding system undergoes a Peierls
instability~\cite{Peierls:55}. While this model is based on the
specific and uniquely simple model of polyacetylene, we believe that
the results we obtain apply more broadly to semiconducting
conjugated polymers. Indeed, similar tight-binding models have
proven useful in predicting not only the properties of
polyacetylene, but have been found to be a useful model for more
chemically complex conjugated polymers such as
PPV~\cite{Fink:86,Yu:95}.  Thus, we present this model as a general
one for the interactions between the electronic degrees of freedom
and the conformational state of the polymer.

Our principal results may be summarized as follows. We find (section
\ref{sec:singlesection}) that the magnitude of the single-site
binding energy and the nature of the bound electron wavefunction are
both very similar for the binding metallic or semiconducting chains.
This similarity is due to the fact that the binding energy is
associated with the creation of localized states having energies far
from the Fermi energy,  so that the density of states at the Fermi
energy in the unperturbed system is not relevant to the inter-chain
interaction energy. These results were discussed previously in
Ref.~\cite{Schmit:08}. We also examine tunneling site interactions
with each other (section \ref{twositesection}) and with solitons
that are necessarily present on sufficiently long dimerized chains
(section \ref{sec:solitons}). We also compare the lowest-energy
bound state configurations of semiconducting and metallic chains.
Although these two systems seem similar at the level of individual
tunneling sites, in section \ref{sec:bindingsymmetry} we find that
the ground states differ rather dramatically~\cite{Schmit:08}. We
also discuss the implications of our numerical quantum chemical
calculations for the stability of these ground states, determined by
the SSH model. In section \ref{sec:polymer} we employ these results
in the development of a modified Poland-Scheraga model to describe
inter-chain binding and the conformational statistics of the result
bound chain pairs. We conclude in section \ref{sec:conclusions}, and
in the Appendices we present the details of our calculations and
numerical quantum chemical calculations supporting our binding
mechanism.

\section{Isolated Binding Sites}
\subsection{A Single Crossing Point}
\label{sec:singlesection}

The binding between conjugated polymers is a generic property of the
mixing of partially filled $\pi$-orbitals. We treat the electronic
component of the interaction using a simple tight-binding model. To
introduce a band gap we use the elegant SSH formalism where a
Peierls instability opens a band gap through a distortion of the one
dimensional lattice. One could introduce similar tight-binding
models with a more complex unit cell appropriate for e.g.
PPV~\cite{Kirova:99}, but such calculations mask the underlying
physics of the system in its most straight forward version. This
simplification of the problem has been exploited previously by Guo
and collaborators, who showed that the SSH model can be used to
study other polymers, such as PPV, with a suitable reinterpretation
of the model parameters~\cite{Guo:95}. Of course, these chemically
distinct systems would produce quantitatively different binding
interactions, but our primary findings, e.g. that there is an
attractive interaction due to the creation of localized states above
and below each band, should be generally insensitive to their
chemical details. The SSH model also allows us to explore in an
analytically tractable manner the interaction between inter-chain
binding sites and local lattice distortions such as solitons. These
results are also generalizable to other conjugated polymer systems
lacking this ground state degeneracy as superpositions of soliton
states can be used to model the more general polaron problem in
these systems~\cite{Heeger:88}.

\begin{figure}[htpb]
\centering
\includegraphics[width=8.0cm]{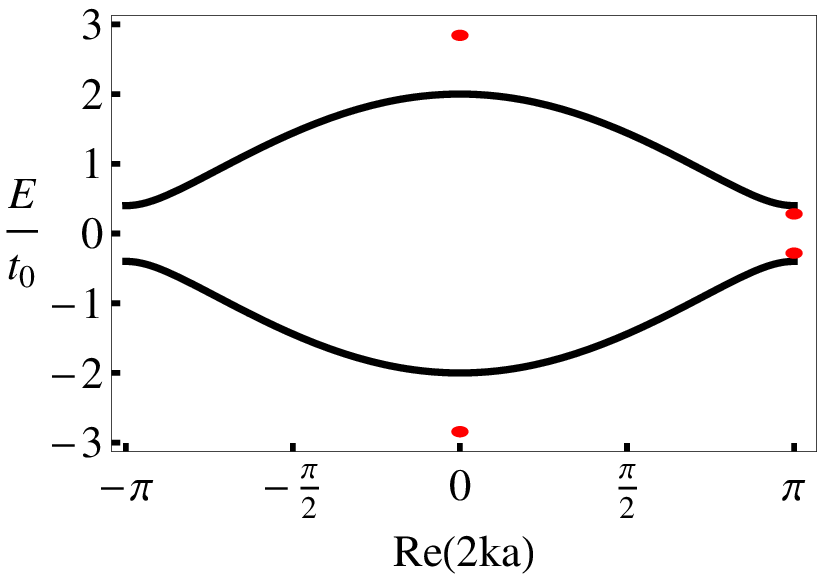}
\includegraphics[width=8.0cm]{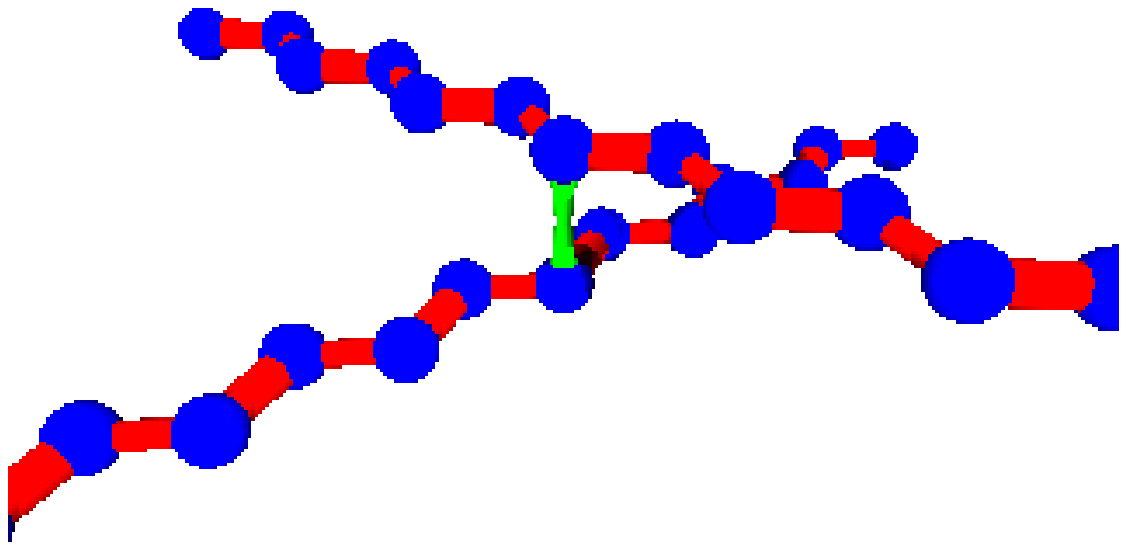}
\includegraphics[width=8.0cm]{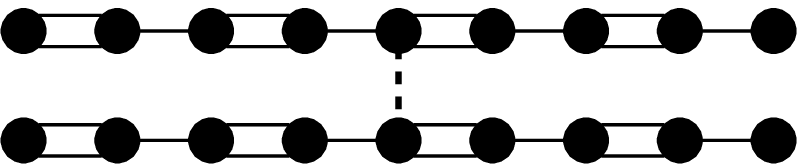}
\caption{(color online)(top) The spectrum of electronic
states from Eq. \ref{eq:dispersion} is plotted against wavenumber $k$, shown as (black) solid lines, along
with the four bound states at $k/2a=0,\pi$, shown as (red) filled circles, resulting from the
single site interaction Eq. \ref{interaction}.
(middle) Cartoon of two polyacetylene molecules
with one tunneling site (green) where there is a significant overlap of
the $p_z$ molecular orbitals. The carbon atoms are shown as (blue)
spheres connected by (red) lines representing the sigma bonds. (bottom)
Schematic of the Hamiltonian that describes this configuration. The
double (=) and single (-) lines correspond to the matrix
elements $t_1$ and $t_2$ respectively.  The dotted line indicates
the inter-chain interaction ($t'$).}
\label{fig:crossmodel}
\end{figure}

The SSH Hamiltonian is given by
\begin{eqnarray}
\label{sshH}
H_{SSH}&=&-\sum_{\ell,\sigma}t_{\ell,\ell+1}(\ket{\ell+1,\sigma}
\bra{\ell,\sigma}+\ket{\ell,\sigma}\bra{\ell+1,\sigma}) \nonumber \\
&&+\sum_\ell\frac{K}{2}(u_{\ell+1}-u_\ell)^2.
\end{eqnarray}
Here and throughout this article we neglect the dynamics of phonons
not associated with the soft phonon mode at the edge of Brillouin
zone. We will not consider electron--phonon scattering, but we will
allow for more complex static lattice distortions when computing the
interaction of tunneling sites and solitons. Here
$\ket{\ell,\sigma}$ is the tight-binding state at the $\ell$th site
on the chain with spin $\sigma$, $K$ is the effective spring
constant of the $\sigma$ bonds, and $u_\ell$ is the displacement of
the $\ell$th carbon from its equilibrium position at $\ell a$.
Introducing the Peierls instability for the system with one electron
per site we generate static lattice distortions $u_\ell =
(-1)^{\ell} \bar{u}$ (on the scale of $\bar{u} \simeq 0.04 \AA$) and
shift the hopping matrix elements so that
$t_{\ell,\ell+1}=t_0-\alpha(u_{\ell+1}-u_\ell)$, doubling the unit
cell of the lattice. Now the electronic states of the system satisfy
the dispersion relation (see
Fig.~\ref{fig:crossmodel}a)
\begin{equation}
E(k)=\pm\sqrt{t_1^2+t_2^2+2t_1 t_2\cos(kb)},\label{eq:dispersion}
\end{equation}
which has a total bandwidth of $4t_0$ and a bandgap centered around
$E=0$ of width $8u_0\alpha\equiv2\Delta$ \cite{Su:79}.

To study the interaction between two chains, we add a second
chain with the same Hamiltonian as Eq. \ref{sshH}
and an interaction term of the form
\begin{equation}
\label{interaction}
H_I=-t'(\ket{0,1,\sigma}\bra{0,2,\sigma}+\ket{0,2,\sigma}\bra{0,1,\sigma}),
\end{equation}
where we have defined the origin to lie at the interaction site, and
we have labeled the chains as 1 and 2.  The inter-chain hopping
parameter $t'$ implicitly includes the effects of the medium and the
relative orientation of the chains.  It is also a function of the
distance between the linked tight-binding sites with a decay length
on the order of a Bohr radius.  Quantum chemical calculations on
similar conjugated systems have shown that $t'$ is on the order of
0.1 eV \cite{Bredas:02}.  We assume the chains have adopted a
conformation where only one pair of sites is close enough for the
inter-chain hopping to be significant. This configuration is
illustrated in Fig.~\ref{fig:crossmodel}b and c.

The interaction part of the Hamiltonian given in
Eq.~\ref{interaction} can be diagonalized using basis of states that
are symmetric or anti-symmetric combinations of the single-chain
states,
\begin{eqnarray}
\ket{\ell,s,\sigma}&=&2^{-1/2}(\ket{\ell,1,\sigma}+\ket{\ell,2,\sigma}) \nonumber \\
\ket{\ell,a,\sigma}&=&2^{-1/2}(\ket{\ell,1,\sigma}-\ket{\ell,2,\sigma})
\label{symmetrize}
\end{eqnarray}
In this basis the Hamiltonian of the system breaks up into two
copies of the single chain Hamiltonian (Eq. \ref{sshH}) that are
identical except for the presence of a diagonal element in each
sub-space of the Hamiltonian at the interaction site with the value
$-t'(t')$ for the symmetric (anti-symmetric) sub-space.

The tunnel matrix element is now analogous to that of an impurity
atom in a one dimensional crystal.  The impurity potential gives
rise to bound states by scattering growing states into decaying
states. In appendix~\ref{apdx:matrix} we use a transfer matrix
technique to show that the interaction leads to the creation of four
bound states as illustrated by the (red) dots in
Fig.~\ref{fig:crossmodel}a. Two of these states,
termed ``ultraband'' states \cite{Phillpot:87}, appear above the
conduction band and below the valance band with energies
\begin{equation}
E_u=\pm\sqrt{t_1^2+t_2^2+\frac{t'^2}{2}+\sqrt{4t_1^2
t_2^2+t'^2(t_1^2 +t_2^2)+\frac{t'^4}{4}}},\label{eq:boundstatesU}
\end{equation}
while the other two appear in the gap and at the edges of the
Brillouin zone with energies
\begin{equation}
E_g=\pm\sqrt{t_1^2+t_2^2+\frac{t'^2}{2}-\sqrt{4t_1^2
t_2^2+t'^2(t_1^2 +t_2^2)+\frac{t'^4}{4}}}.\label{eq:boundstatesG}
\end{equation}
If we assume that $t', \Delta \ll t_0$ we find that, to lowest
order, the ultraband states have been shifted away from the band
edge by $t'^2/4t_0$ which is unchanged from the metallic case
\cite{Schmit:05}. In contrast, the gap states are shifted by
$t'^2\Delta/8t_0^2$.  For half-filled chains, only the two bound
states associated with the valence band will be filled.  Because the
shift of the gap state is smaller than the shift of the ultraband
state by $O(\Delta/t_0)$ for typical parameter values, the result is
a net lowering of the energy.

The presence of the binding site will also perturb the energy of the
extended electron states.  However, as in the metallic case, it can
be shown that the total contribution of these states is $O(N^{-1})$,
where $N$ is the total number of tight-binding sites on the polymer,
and therefore can be neglected for long chains \cite{Schmit:05}.

In appendix \ref{sec:hartree} we present Hartree-Fock (HF)
calculations that verify the qualitative features of this binding
mechanism.  These calculations show that the binding energy is of
order $k_BT$ and that it is sufficiently insensitive to the precise
orientation of the polymers to allow for a variety of aggregate
morphologies.

\subsection{Two Binding Sites}
\label{twositesection}

Since conjugated polymers are semiflexible,  it is possible for them
to bind in multiple places separated by arc lengths of unbound
chains.  Two such distant binding sites will experience an effective
interaction due to a combination of the change in chain
configurational entropy due to the binding constraint, the bending
elasticity of the chains, and possibility the electrostatic
repulsion of the backbones~\cite{Borukhov:01}. In addition, there
may be a modification of the binding energy of each binding site due
to the interaction of the electronic wavefunctions associated with
each of them. We now study this latter effect by determining the
shift in total bound state energy of two binding sites as a function
of their separation along the chain.

Previously, we demonstrated that the energy shift of an extended
state in the presence of a single impurity could be expressed as a
series in $t'/t_0 N$ \cite{Schmit:05}.  Because the symmetric and
anti-symmetric states differ only in the sign of the impurity, the
odd terms in this series will cancel.  Therefore, after summing over
the ${\cal O}(N)$ extended states, the total energy contribution
from these states must vanish as $N^{-1}$.

\begin{figure}[htpb]
\centering
\includegraphics[width=8.0cm]{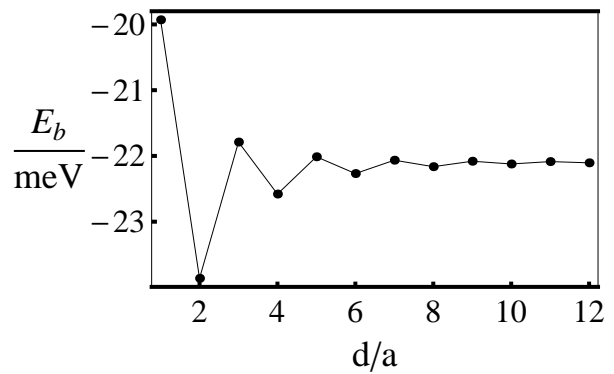}
\caption{The interaction potential of two tunneling sites as a
function of the distance $d$ between them measured in units of the
unperturbed lattice $a$, as determined by the
numerical diagonalization of chains 100 tight binding sites each.
The tunneling matrix elements are given by $t'=0.25$ eV. For these
parameters, the energy difference between bound $d/a=2$ and free
$d/a > 10$ tunneling sites is approximately $0.1 k_BT$. This weak
attraction increases with $t'$. Entropic effects, however, play a
comparable role in binding site interactions.} \label{fig:twositesD}
\end{figure}

A second binding site introduces a new length $d$, the distance
between the binding sites. We expect the perturbation series for the
extended states to contain terms proportional to $d^{-1}$,
suggesting that the contribution of the extended states may be
non-vanishing for long chains. This is indeed the case. Taking these
effects into account, we determine the interaction energy of two
binding sites by direct numerical diagonalization of
Eqs.~\ref{eq:transfer}; the result is shown in
Fig.~\ref{fig:twositesD}. There is an attractive potential well for
the two tunneling sites with a minimum at the separation of $2 a$.
At separations greater than about 10 tight binding sites the binding
energy is essentially constant, however, at all smaller separations
there is a net reduction in energy for even separations and a
corresponding increase in energy for odd separations.

This ``even-odd'' effect can be understood by looking at the states
that receive the maximum perturbation from the binding potential.
From first order perturbation theory, we expect that the energy
shift of a given state to be proportional to the amplitude of that
state at the two impurities.  Therefore, the states perturbed the
most by the two impurities separated by distance $d$ will be those
with a dominant wavenumber $k$ that satisfies the relation
$kd=n\pi$.  However, the interaction potential induces a phase
shift, $\phi$ in the wavefunction.  So, the condition for the
maximally perturbed state is actually $kd+\phi=n\pi$. Because $\phi$
is positive for attractive impurity potentials and negative for
repulsive potentials~\cite{phaseshift}, the symmetric states will
see the full effect of the impurity potential at smaller values of
$k$ than the anti-symmetric states.

For the present case of half-filled electronic states, the highest
occupied level has a wavenumber $k_f=\pi/2a$.  For even $d/a$ we
find that $k_f d=n\pi$ so the states with large negative shifts are
filled but the corresponding states with large positive shifts are
unoccupied. However, if $d/a$ is odd the states with large positive
shifts are also filled resulting in an increase in energy relative
to the two isolated binding sites. The interaction between two
binding sites is discussed in more detail in
Appendix~\ref{apdx:siteinteract}.  Given these results, we expect
that binding sites should have a weak tendency to cluster so that
chains develop finite bound sections in thermal equilibrium. Any
conclusions regarding bound-state chain configurations will have to
be postponed until we consider the chain configurational entropy of
these bound states. We turn to this problem in
section~\ref{sec:polymer}. Before we do so, we consider the problem
of multiple inter-chain binding sites (instead of just two) as might
occur when the chains adopt a parallel alignment.

\section{Binding/Non-binding Symmetry}
\label{sec:bindingsymmetry}

\begin{figure}[htpb]
\centering
\includegraphics[width=8.0cm]{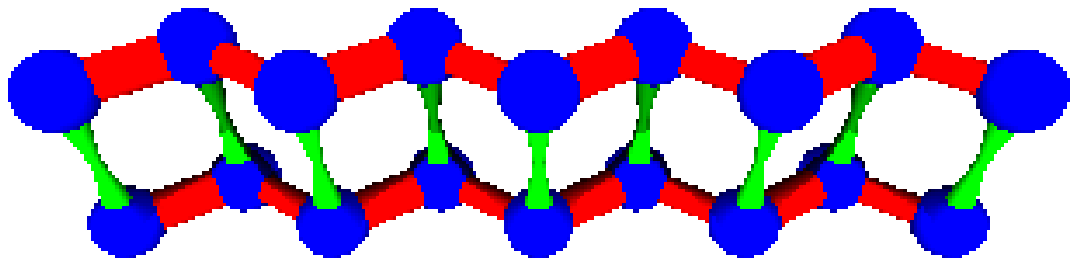}
\includegraphics[width=8.0cm]{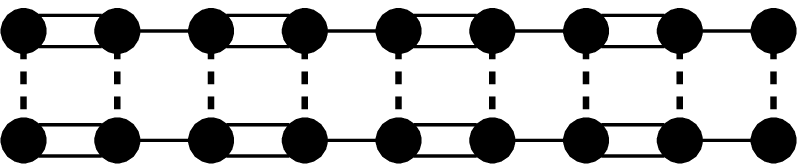}
\includegraphics[width=8.0cm]{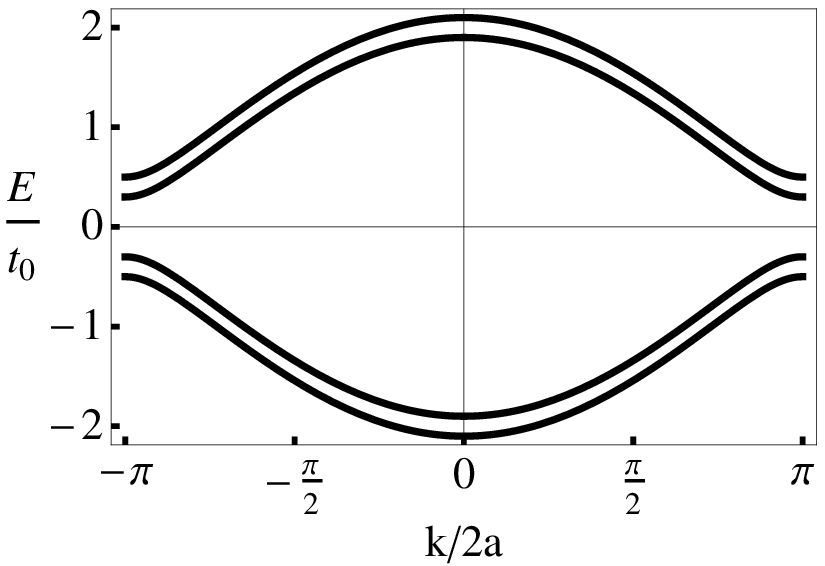}
\caption{(color online)(top) Cartoon of the paired chains having
tunneling sites at every tight binding site. The color scheme is
identical to that of Fig.~\ref{fig:crossmodel}.  (middle) Schematic
representation of the Hamiltonian describing same set of tunneling
sites. (bottom) The electronic band structure for the paired chains
with tunneling at every site.} \label{fig:fullzipfig}
\end{figure}

If two chains are aligned in parallel, it is possible to have
binding events at every site along the chain as illustrated in
Fig.~\ref{fig:fullzipfig}.  The interaction Hamiltonian is then
\begin{equation}
H_I=-t'\sum_\ell(\ket{\ell,1,\sigma}\bra{\ell,2,\sigma}+
\ket{\ell,2,\sigma}\bra{\ell,1,\sigma}).\label{eq:fullinteract}
\end{equation}
This interaction can also be diagonalized using the change of basis
shown in Eq.~\ref{symmetrize}. We find that both the valence and
conduction bands are split with the chain-symmetrized states
dropping in energy by $t'$, and the chain-anti-symmetrized ones
rising by $t'$; see Fig.~\ref{fig:fullzipfig}. For conjugated
polymers interacting in solution, however, $t'$ is considerably
smaller than $\Delta$.  Thus, the symmetric and anti-symmetric bands
have equal occupancy, and there is no net change in energy and no
binding.

At first it seems counter-intuitive that the chains can bind at a
single site, but have no attraction with many binding sites. To
understand this result we now show that there is a symmetry between
chains having spatially uniform densities of binding sites $n$ and
$1-n$. The lack of binding at maximal tunneling site density $n=1$
is thus understandable since this state is energetically equivalent
to no binding sites at all:  $n=0$.  This symmetry
results in a maximally bound state with interchain tunneling at
every other site, i.e. $n=1/2$~\cite{Schmit:08}.

To observe this proposed symmetry, consider first the Hamiltonian of
chains bound at every site: $H_{FB}=H_0+H_I'$ where $H_0$ is the
symmetrized/anti-symmetrized non-interacting Hamiltonian, and $H_I'$
is the symmetrized version of Eq.~\ref{eq:fullinteract}.   As shown
in Fig.~\ref{fig:fullzipfig}c, this Hamiltonian is identical to
$H_0$ apart from equal and opposite constant shifts in the symmetric
and anti-symmetric bands. If we now remove the overlap at a single
site, we have effectively introduced an (non-binding) impurity site
having a potential of strength $+t'(-t')$ in the symmetric
(anti-symmetric) band. Because of the symmetry $t' \rightarrow -t'$
in Eqs.~\ref{eq:boundstatesU} and \ref{eq:boundstatesG}, the removal
of an interaction at a single site results in the same net energy
change as adding a single interaction site to the non-interacting
chains. This argument can be extended to any set of interacting
sites. Allowing a set $\{\alpha\}$ of sites to interact on the
unbound chains will change the energy of the symmetric band by the
same amount as the energy change of the anti-symmetric band upon the
removal of interactions from the same set $\{\alpha\}$ from the
fully bound chains. Similarly, the anti-symmetric band on the
non-interacting chains changes by the same energy as the symmetric
band on the fully bound chains upon the addition/subtraction of
interactions at $\{\alpha\}$.
\begin{figure}[htpb]
\centering
\includegraphics[width=8.0cm]{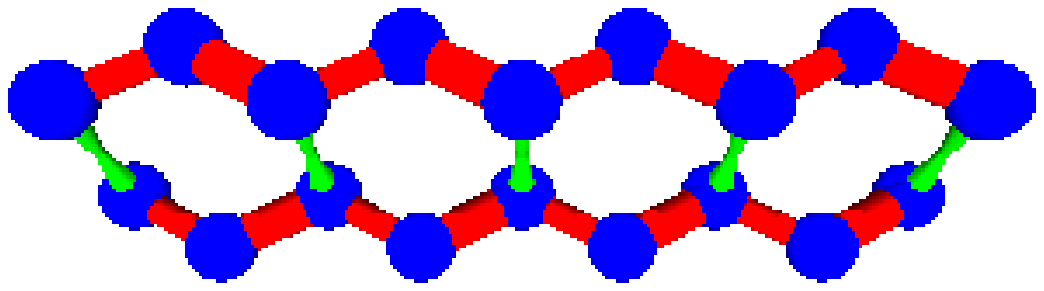}
\includegraphics[width=8.0cm]{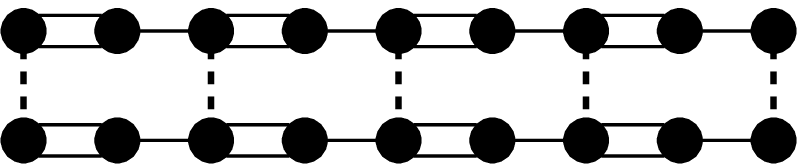}
\caption{(color online)(top) Cartoon of the paired chains having
tunneling sites at every other tight binding site, using the color
scheme of Fig.~\ref{fig:crossmodel}.  This is the tunneling
configuration that results in the lowest electronic energy of the
pair by increasing the Peierls gap. (bottom) Schematic of the
Hamiltonian describing this configuration.} \label{fig:halfzipfig}
\end{figure}

The electronic ground state of the pair of parallel chains, as a
consequence of this symmetry, requires tunneling at every-other
site. The all-trans configuration of the carbon backbone of the
polyacetylene molecule as shown in Fig.~\ref{fig:halfzipfig} allows
for precisely the required geometry. It is interesting to note that
this tunneling site pattern effectively widens the Peierls gap.

More complex conjugated polymers may not be able to achieve this
maximally bound state due to e.g. steric interactions of side
chains.  In spite of the symmetry between pairs of chains having
binding density $n$ and $1-n$, we do not expect to observe cases
where $n>1/2$ as a reduction of the number of inter-chain tunneling
sites will simultaneously decrease the electronic contribution to
the energy of the system and increase the its configurational
entropy.

\section{Solitons}
\label{sec:solitons}

The dimerization pattern of the tight-binding sites in the SSH model
breaks a two-fold symmetry of the system. Reading from the left to
the right the slightly longer bond ($-$) can either precede or
follow the slightly shorter one ($=$). In thermal equilibrium this
one-dimensional system cannot maintain long-range order by keeping
only one of these two dimerization patterns. One expects to find
domain walls where the dimerization pattern switches from
$\circ=\circ-\circ=\circ$ to $\circ-\circ=\circ-\circ$. At the
domain wall, or soliton, the pattern of bond lengths is necessarily
distorted over some finite distance. These solitonic structures have
been studied in detail by Su and collaborators\cite{Su:79}: They
found that there are new localized electronic states associated with
these domain walls. It remains to be seen how these localized states
associated with the solitons are affected by inter-chain tunneling.

For computational simplicity, we first consider the symmetric case
in which the two chains each have a single soliton located at the
tunneling site. This allows us to separate the binding and
anti-binding states using Eq.~\ref{symmetrize}. Furthermore, we
simplify the structure of the soliton~\cite{Su:79} to that a
single-site domain wall. Using our notational short hand, this
soliton can be represented as \linebreak
$\circ=\circ-\circ=\circ-\bullet-\circ=\circ-\circ=\circ$ where the
filled circle represents both the center of the soliton and the
location of the binding site.  In Appendix~\ref{apdx:matrix} we show
that the bound states associated with this scattering center are
given by the solutions of cubic polynomial
\begin{equation}
0=E^3-E(4t_0^2+\Delta^2+t'^2)\pm 4t' t_0\Delta.
\label{eq:solitondispersion}
\end{equation}
In the limit that $E\gg t',\Delta$ we find roots corresponding to
ultraband bound states at energies
$\pm(4t_0^2+\Delta^2+t'^2)^{1/2}$, which, except for the $\Delta^2$
term, are identical to the result for metallic
chains~\cite{Schmit:05}.  We also find another root of
Eq.~\ref{eq:solitondispersion} in the gap. This mid-gap state has an
energy of $4t' t_0 \Delta/(4t_0^2+\Delta^2+t'^2)$.  The presence of
two solitons has minimal effect on the ultraband bound states, while
the mid-gap states associated with the solitons are split by ${\cal
O}(t' \Delta/t_0)$. For half-filled chains the lower gap state is
filled for both spin states, while the corresponding upper state
remains empty in the electronic ground state of the system. Although
solitonic perturbation is linear in $t'$, as opposed to the
ultraband energy shifts which are ${\cal O}(t'^2)$, the net energy
change is comparable to the binding energy from ultraband states due
to the additional factor of $\Delta/t_0$. The net result is an
${\cal O}(1)$ enhancement to the binding energy due to the
co-localization of the solitons at the inter-chain tunneling site.
It is important to note that the binding energy enhancement
associated with solitonic co-localization occurs only for uncharged
solitons, since charged ones contribute either empty or doubly
occupied mid-gap states. The splitting of the filled mid-gap states
by the tunneling matrix element results in no net energy change of
the system. Now, the energy required to form a soliton is
approximately $0.42$eV~\cite{Su:79} suggesting the thermal
equilibrium density of uncharged solitons is typically small, but
they can also be trapped on long chains by the process of cis- to
trans-isomerization; additionally a single uncharged soliton will
spontaneously form on chains with an odd number of
sites~\cite{Su:80a,Su:80b}. Thus, soliton co-localization can play a
role in the strengthening of isolated inter-chain binding sites, but
presumably is less relevant as the number of binding sites
increases.

We studied the co-localization of uncharged solitons with the
tunneling site numerically in order to explore a more physical
extended soliton. Using a $\tanh[s/(\Delta s)]$ profile for the
static displacement field of the soliton, we verified the ${\cal
O}(1)$ enhancement of the binding energy. The calculation was
performed on two chains of 199 sites each with a
soliton centered at site 101. We also
varied the width of the soliton $\Delta s$ to ascertain if the
presence of the tunneling site altered the minimum energy structure
of the soliton as determined by Su and collaborators~\cite{Su:79}.
We found that such an effect was negligible.

\begin{figure}
\includegraphics[width=8.0cm]{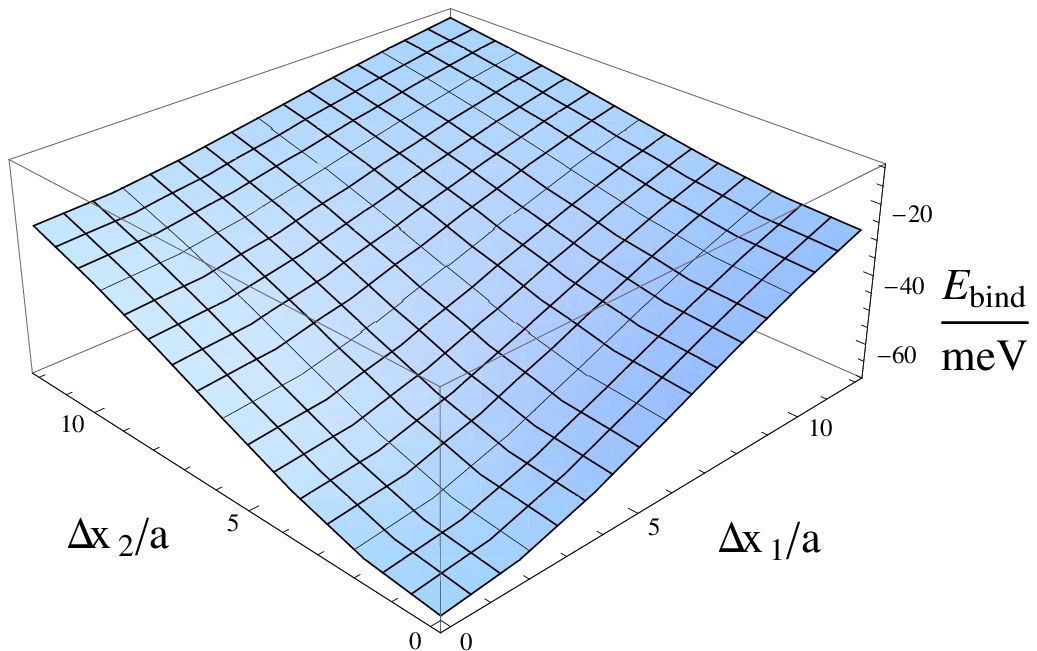}
\includegraphics[width=8.0cm]{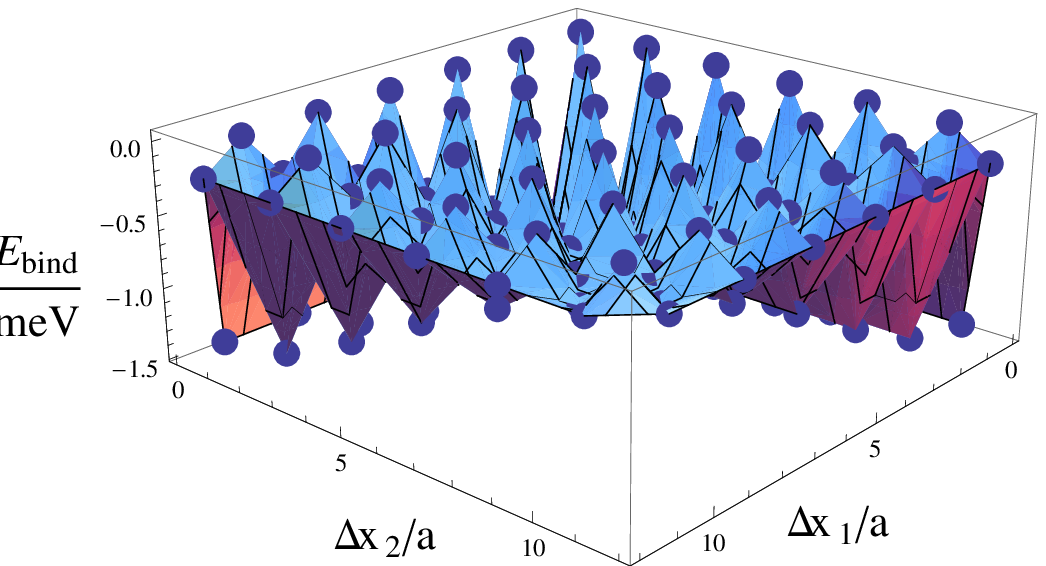}
\includegraphics[width=6.0cm]{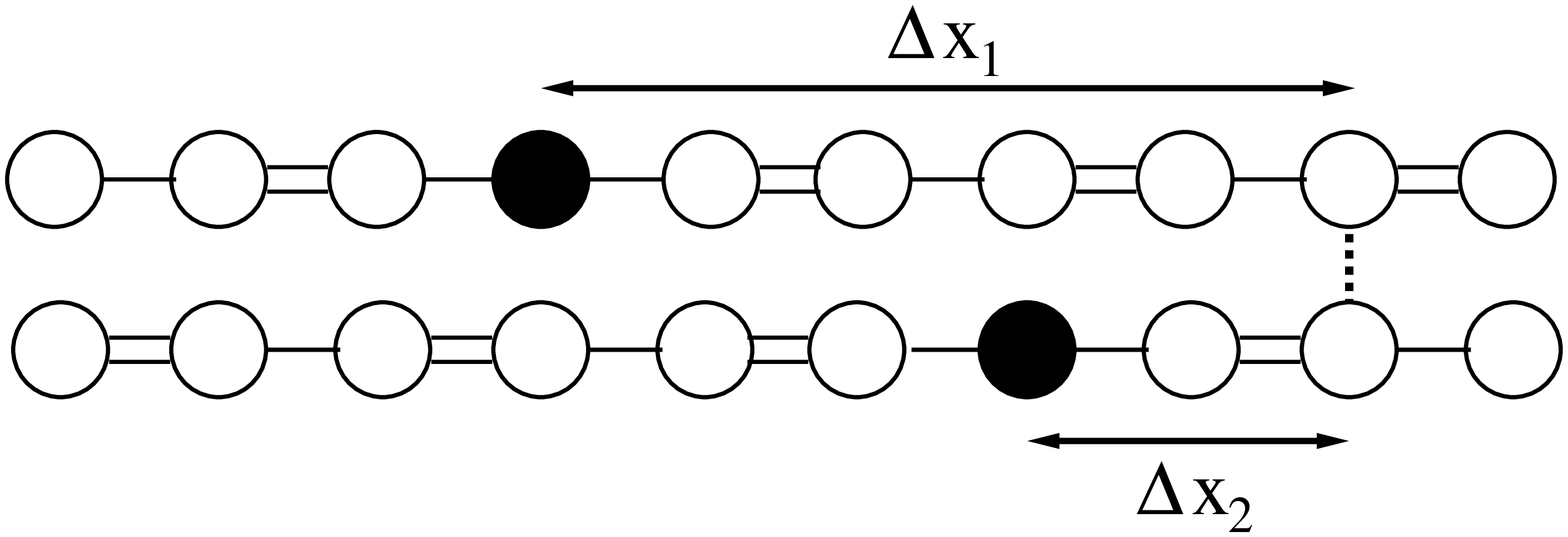}
\caption{a) (color online) The single tunneling site binding energy
of two chains (labeled (1) and (2)), each having a soliton centered
at a distance $\Delta x_1$, $\Delta x_2$ from the tunneling site.
The minimum energy (maximum binding) occurs when both solitons are
localized at the tunneling site: $\Delta x_1 = \Delta x_2 = 0$. Due
to the ``even-odd'' effect mentioned in the text, we represent the
results as two surfaces. (top) The binding energy
for cases where the distances $\Delta x_1,\Delta x_2$ represent even numbers of interatomic distances.
(middle) The corrugated energy surface showing the binding energy for separations
where at least one of $\Delta x_1$ or $\Delta x_2$ are odd.  Note
the change in energy scale from above.  (bottom)  A schematic
representation of the centers of solitons at distances $\Delta x_1$
and $\Delta x_2$ from the tunneling site (dotted line).
The figures were calculated using $N=199$, $4t_0=10$ eV,
$u_0=0.04$ \AA, $K=21\ {\rm eV/\AA}^2$, $\alpha=4.1$ eV/\AA,
$t'=0.25$ eV, and $\Delta s=7a$.}
\label{fig:soliton-binding}
\end{figure}

More generally, we determined the change in electronic ground state
energy of two chains forming one tunnel junction and each having one
soliton whose center is located $\Delta x_{i}$ sites from the
tunneling site. Here the index $i$ labels the chains. These results
are shown in Fig.~\ref{fig:soliton-binding}. The principal result is
that both solitons are strongly attracted to the the tunneling site,
but the strength of the attractive potential of a soliton on one
chain is significantly enhanced when the soliton on the other chain
is already localized at the tunneling site.  This effect results
from the mixing of the mid-gap states on each chain due to the
inter-chain tunneling matrix element and leads to a further
reduction of the energy of the filled mid-gap states.
We also note the large amplitude oscillation of the solitonic interaction energy with
distance from the tunneling site. This is simply due to the fact
that the wavefunction of the mid-gap states has nodes on alternating
tight-binding sites.

\section{Polymer Degrees of Freedom}
\label{sec:polymer}

For polymers in solution, the morphology of their aggregates is
determined by the relative strength of the intermolecular binding
mechanism and the conformational degrees of freedom of the polymers.
These parameters will depend strongly on variables like
electrostatic screening and solvent mediation of the interchain
hopping parameter.  Since these are expected to vary greatly
system-by-system, we present a general calculation describing the
onset of aggregation.

In sufficiently dilute solution,  we expect the formation of stable
two-polymer aggregates. These may take the form of loosely-bound,
braid-like structures in which the polymers form non-interacting
loops between rare binding sites, or zipper-like with the polymers
interacting at many consecutive sites~\cite{Borukhov:01}. Since the
single-site binding mechanism discussed in
section~\ref{sec:singlesection} is qualitatively unchanged from the
doped case, the braid structures will also be unchanged from
conducting polymers discussed earlier~\cite{Schmit:05}. The
formation of the tightly-bound, so-called ``zipper structures,''
however, differs between the conducting and semiconducting polymers.
As shown in that previous work on the doped, metallic system, the
length of the tightly bound regions is limited by the doping level
of the polymers. In the case of undoped, semiconducting polymers,
the binding energy due to long, tightly-bound regions grows linearly
with their length, provided the binding sites occur at every other
site as shown in Fig.~\ref{fig:halfzipfig}. For a pair of bound
polymers interacting via a series of well-separated tightly-bound
regions, the total density of states will be equal to the sum of the
density of states for independent segments of tightly bound and free
chains weighted by the respective lengths. For this situation (with
binding sites at every other site) the dispersion relations for the
conduction bands are given by
\begin{equation}
E(k)=\pm\frac{t'}{2} +
\sqrt{\left(\frac{t'}{2}\right)^2+t_1^2+t_2^2+2 t_1 t_2 \cos(2 k
a)};
\end{equation}
the valance bands have equal magnitude and opposite signs. We find
that the binding energy per site in the tightly bound segments is
then given by
\begin{eqnarray}
E_z&=&\frac{-8 a}{\pi}\int_0^{\pi/2 a} \left[
\sqrt{\left(t'/2\right)^2+t_1^2+t_2^2+2 t_1 t_2
\cos(2 k a)} \right .  \nonumber \\
&&\left . -\sqrt{t_1^2+t_2^2+2 t_1 t_2 \cos(2 k a)} \right ] dk \\
&\simeq&\frac{-t'^2}{2t_0 \pi}\int_0^{\pi/2} d\theta
\left(\cos^2\theta +\left(\frac{\Delta}{2t_0}\right)^2\sin^2
\theta\right)^{-1/2} , \label{zipbind}
\end{eqnarray}
where $\theta=k a$ and additional factors of two
have been included to account for spin, positive and negative values
of $k$, and the fact that two chains are present.
This binding energy is the product of the same dimensionful prefactor  $t'^2/2t_0$,
found for the single site binding energy, and a dimensionless
integral that depends only on the underlying lattice.  For
parameters typical of polyacetylene this integral generates a factor of $\sim 1.1$,
indicating a modest cooperative effect. However, the integral in Eq.
\ref{zipbind} is sensitive to the ratio $\Delta/2t_0$ and the
cooperativity may be significant in other polymer systems.  The
transition from single site binding behavior to zipper-like binding
occurs when the bound region grows comparable to the localization
length $a t_0/t'$.

Without the doping level to limit the length of the zipped region,
the amount of chain in the tightly-bound and free states is
determined by a competition between the decrease in the system's
total electronic energy and the decrease in the chains'
configurational entropy associated with the binding regions. The
statistical mechanics of the binding of two conjugated polymers is
thus similar to that of DNA melting~\cite{Poland:70}. Here we adapt
the Poland-Scheraga (PS) model designed for that problem to the
current context. There are, however, significant differences between
the two systems arising from the difference in bonding mechanism and
the lack of unique binding sites on the conjugated polymers, due to
the absence of the well-known DNA base pairing mechanism in the
system of current interest.

A pair of bound polymers consists of alternating regions of
``zipped'' (i.e. tightly bound) chains separated by loops of unbound
polymer. The partition function for chains of length $N$ in such a
configuration may be written as
\begin{equation}
Z(N)=\sum_{p}\sum_{\{i_\pi^{(1)},i_\pi^{(2)},j_\pi\}}\prod_{\pi=0}^p
u(i_\pi^{(1)},i_\pi^{(2)})v(j_\pi).
\label{PSpartition}
\end{equation}
Here $u(i_\pi^{(1)},i_\pi^{(2)})$ and $v(j_\pi)$ are the Boltzmann
weights of the $\pi^{\rm th}$ loop and the tightly bound domain
respectively, enumerated from the left end of each chain. The number
of sites in a tightly bound or ``zipped'' region, $j_\pi$, must be
an even number as each chain contributes $j_\pi/2$ sites. The size
of the loop regions, $i_\pi^{(1)}+i_\pi^{(2)}$, are unconstrained
because the translational symmetry of the chains allows the
formation of asymmetric loops (i.e. the number of sites contributed
by chain 1, $i_\pi^{(1)}$, is not necessarily equal to the number of
sites that chain 2 contributes to the loop, $i_\pi^{(2)}$). The sums
over the number of ``zipped'' and unbound regions $p$ and the length
of each region are subject to the constraint that the overall
lengths of the chains are fixed
\begin{equation}
\label{Nconstraint} \sum_{\pi=0}^p(i_\pi^{(n)}+j_\pi/2)=N.
\end{equation}

The statistical weight of a ``zipped'' region of $n$ bound sites is
given by $v(n)=v^n \sigma(n)$, where $\sigma(n)$ accounts for the
potentially length-dependent boundary energy between looped and
zipped regions, and
\begin{equation}
\label{v-definition} v= e^{-(E_{zip}-E_{ES})/k_BT}
\end{equation}
is the Boltzmann weight associated with one bound pair of sites (one
from each chain) in the ``zipped'' region.  The energy that appears
in the exponent has been broken into two parts. The first term
$E_{zip}$ is the per site inter-chain binding energy given by
Eq.~\ref{zipbind} for long tightly bound regions and $-2(4
t_0^2+t'^2)^{1/2}+4t_0$ for a single binding site. In practice we
use the form given by Eq.~\ref{zipbind} for ``zipped'' regions of
length greater than $a t_0/t'$. Otherwise, we use the result for an
isolated binding site.  The second term represents a local per-site
repulsion between the chains in close proximity. This energy arises
from unfavorable steric interactions between the chains enforced by
the local binding geometry and possibly electrostatic repulsion
between the chains. The details of this energetic term will clearly
vary with the chemical details of the specific conjugated polymer
system in question. Since $E_{ES}$ is of order $k_BT$, we expect
that the net binding energy may be smoothly varied from a minimum of
$E_{zip}$ to net repulsive values.  The parallel configuration
considered below may also contain a van der Waals component
enhancing the net attraction \cite{Schmit:05}.  However, this
interaction decays much more slowly with distance than interchain
hopping, and therefore the details of aggregation on the monomer
level will be dominated by the mechanism considered here.  We do not
consider these details further here.

The transition between a loop and a zipped region of length $n$
results in a boundary free energy $- k_B T
\ln\left(\sigma(n)\right)$. This boundary energy has contributions
from the cooperativity of the intermolecular bonds, the
configurational entropy cost associated with constraining the
polymers' backbone with the formation of the second bond, and a
bending energy associated with forming the ``Y''-junction where the
collinear polymers split into an unbound loop. In the case of
conjugated polymers, the first two contributions are negligible due
to the weak cooperativity of the bonds and the limited flexibility
of the chains.  This is in contrast to the situation in nucleic
acids where the aromatic base stacking interactions leads to strong
cooperativity in the inter-chain bonds while the flexible backbone
results in a negligible bending energy and a non-negligible entropic
contribution.

For conjugated polymers the statistical weight for the bound regions
takes the form
\begin{equation}
\begin{array}{lll}
v(n)&=0 & n={\rm odd} \\
&=v & n=2 \\
&=\sigma v^n & n\geq 4.
\label{Vdef}
\end{array}
\end{equation}
This form accounts for the fact that if the ``zipped'' region is
only a single site long, i.e. a single binding site, the angle
between the chains is only weakly constrained.  For longer tightly
bound regions, however, the chains must be parallel over its length
then bend sharply at the start of the surrounding loops. The
boundary parameter $\sigma$ may be estimated using the conjugation
length in a melt. Under these conditions the polymer has conjugation
breaks approximately every ten sites \cite{Heun:93}. Since there
must be two such breaks per tightly bound region and there are four
ways of arranging the breaks, the overall Boltzmann weight should be
$\sigma \sim 0.04$.

The statistical weight for an unbound loop of the paired chains
accounts for the configurational entropy of the self-avoiding loop
when the loop's arc length is long compared to the thermal
persistence length of the polymer. When the loop is short compared
to this length, the bending energy of the chain makes the dominant
contribution to the free energy of the loop. These two effects are
captured using a weight of the form
\begin{equation}
u(n)=e^{- \pi \ell_p/n} u^n (n/2)^{-c},
\label{Udef}
\end{equation}
with $u=e^{2 a/\ell_k}$. Here $\ell_k$ is a length on the order of
the persistence length $\ell_k \sim \ell_p \gg a$, and we have
assumed that short loops trace out the arc of a
circle~\cite{Schmit:05}. The constant $c$ accounts for the excluded
volume of the chains and takes the value $\sim 2.1$ in three
dimensions~\cite{Kafri:00}.

The sums in Eq.~\ref{PSpartition} are difficult to evaluate because
of the restriction imposed by Eq.~\ref{Nconstraint}. They can be
made more tractable by relaxing this constraint. In order to do so
we study the function
\begin{equation}
\Gamma (x)=\sum_{N=1}^\infty \frac{Z(N)}{x^N},
\label{gammasum1}
\end{equation}
which amounts to working in the grand canonical ensemble of the
polymer length where we have introduced a fugacity $x^{-1}$. With
the use of Eq.~\ref{PSpartition} this sum can be rewritten as
\begin{eqnarray}
\Gamma (x)&=&L(x)R(x)\sum_{p=0}^\infty \left( U(x)V(x)
\right)^p \nonumber \\
&=& \frac{L(x)R(x)}{1-U(x)V(x)}, \label{gammasum2}
\end{eqnarray}
where the functions $U$ and $V$ are defined by
\begin{eqnarray}
U(x)&=&\sum_{n=1}^\infty \frac{u(n)}{x^N} \nonumber\\
V(x)&=&\sum_{n=1}^\infty \frac{v(n)}{x^N}. \label{UVdef}
\end{eqnarray}
The functions $L(x)$ and $R(x)$ are defined in analogy to
Eqs.~\ref{UVdef} and account for the free energy of the tails at the
ends of the chain.  $L(x)$ and $R(x)$ will have a different
functional form than the portions of chain on the interior of the
complex due to the fact that the chain may terminate in either a
bound ``zipper" or a pair of free tails that are not subject to a
loop closure constraint.  However, these details are unimportant as
the free energy of the system is dominated by the chain interior in
the thermodynamic limit.

In the limit of an infinitely long pair of chains i.e. where $N
\longrightarrow \infty$, the partition function $Z(N)$ of the
two-chain system must scale as $x_1^{2N}$, so that the free energy
per site is now independent of chain length. The parameter $x_1$ is
as yet undetermined, but is related to the free energy per monomer
$f$ by $f = -k_BT \ln (x_1)$. From this we recognize the sum shown
in Eq.~\ref{gammasum1} will converge for all $x>x_1$. We now
determine $x_1$ by examining convergence properties of this sum as
given by Eq.~\ref{gammasum2}. Noting that the end effects $L(x)$ and
$R(x)$ necessarily generate bounded corrections, we see that the
divergence of the sum will occur at the roots of $U(x)V(x)=1$; we
will ignore all such end effects here and in the following. As we
want the dominant term in the limit of long polymers, the free
energy is controlled by $x_1$, the \emph{largest root} of this
equation~\cite{Lifson:64}. From Eqs.~\ref{Vdef} and \ref{Udef}, we
find that $x_1$ is then given by the solution of
\begin{equation}
\frac{x(x-v)}{xv-v^2(1-\sigma)}=\sum_{n=1}^\infty e^{- \pi \ell_p/n}
\left(\frac{u}{x}\right)^n \left(\frac{n}{2}\right)^{-c}.
\label{xequation}
\end{equation}
This result allows us to determine the free energy and all
thermodynamic properties of the paired chains.

From this result we may compute any number of physically measurable
quantities involving the structure of paired conjugated polymers. We
will consider three here.  First, we compute the fraction of paired
chains as a function of polymer concentration in dilute solution,
where we will show that, in the limit of sufficiently long polymers,
there is a sharp cross-over between free chains and bound pairs as a
function of concentration. One may imagine that in the strongly
bound limit, more complex bundles of chains should form. Second, we
examine the distribution of free loops in the paired chains as a
function of the strength of the binding interaction. This result may
have measurable consequences for the diamagnetic susceptibility of
conjugated polymer solutions. Finally, we compute using a minimal
set of assumptions the effective persistence length of the paired
chains as a function of the strength of the binding interaction.
Such results have implications for the interpretation of scattering
data from these polymers in solution.

\begin{figure}[htpb]
\centering
\includegraphics[width=8.0cm]{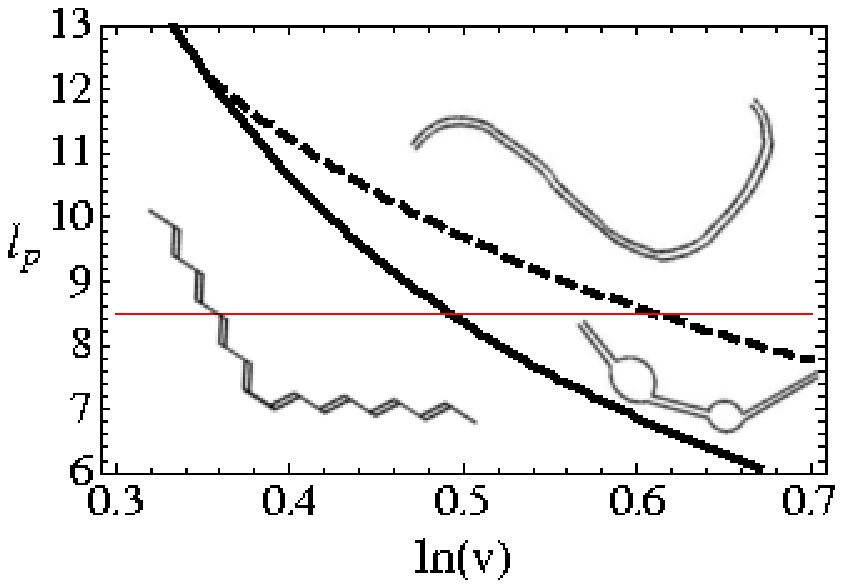}
\includegraphics[width=8.0cm]{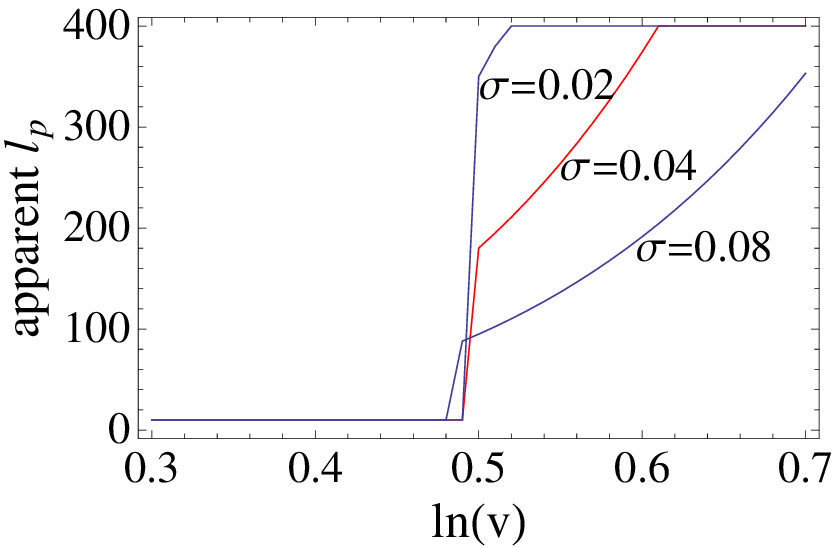}
\caption{ (color online) (top) The apparent persistence length as a
function of interchain binding energy. For sufficiently small
binding energy and for highly flexible (low persistence length)
individual chains, the polymers are remain unbound and have an
apparent persistence length $\ell_{min}$ equal to that of the
individual chains.  As the binding energy increases the polymers
begin to form complexes, indicated here by the solid line which
shows where exactly half of the chains are paired: $2c_2/c_0=0.5$.
For strong enough binding energy, the average length of the zipped
regions $\langle L_{zip}\rangle$ exceeds $\ell_{max}$ (dotted line),
taken here to be $400a$. Between the two lines the apparent
persistence length is given by $\langle L_{zip}\rangle$ since the
paired chains will be essentially straight in their zipped regions,
but have free hinges associated with the looped ones. (bottom)
Apparent persistence length plotted along the red line above (red).
Also shown are curves for smaller and larger values of the
nucleation parameter $\sigma$.  Although we show the persistence
length saturating above $\ln v\simeq0.6$, we expect that
sufficiently strong binding will drive the formation of larger
aggregates, further increasing the apparent polymer rigidity.}
\label{fig:concfig}
\end{figure}

To calculate the fraction of bound chains we write the free energy
for a dilute polymer solution
\begin{eqnarray}
\frac{F}{{\cal V}k_BT}&=&c_1 \left ( \ln \left(c_1\right)-1\right) -
c_1 N
\ln u \nonumber \\
&+&c_2 \left ( \ln \left(c_2\right)-1\right)- c_2 N \ln x_1
\label{eq:solutionFE}
\end{eqnarray}
in terms of the number densities of the unpaired and paired
polymers, $c_1$ and $c_2$ respectively. ${\cal V}$ is the total
volume of the system. To find the fraction of bound chains as a
funtion of the total polymer number density $c_0$ we use the
conservation of the number of chains to write $c_1=c_0-2c_2$ and
minimize Eq.~\ref{eq:solutionFE} with respect to $c_2$. From this we
find that
\begin{equation}
\frac{2c_2}{c_0}=
1+\frac{1}{4c_0}\left(\frac{u}{x_1}\right)^{2N}-
\sqrt{\left(1+\frac{1}{4c_0}\left(\frac{u}{x_1}\right)^{2N}\right)^2
-1}.
\end{equation}
Because of the large polymerization index $N \gg 1$, the transition
from single chains to bound pairs occurs over a very narrow range of
binding energies and becomes even sharper as $N$
increases~\cite{Schmit:08}. The binding curve is shown in
Fig.~\ref{fig:concfig}.

The physical properties of bound chains will be determined in part
by the statistics of the zipped and loop regions. For example, the
magnetic susceptibility of polymeric solution should depend in part
on the number density of semiconducting or conducting (for suitably
doped chains) loops in solution, which itself will depend on the
loop fraction of the individual bound chains.  The fraction of
zipped regions, which are significantly stiffer mechanically, will
determine the effective persistence length of the polymers.  We
first calculate the distribution of unbound loops.

\begin{figure}[htpb]
\centering
\includegraphics[width=8.0cm]{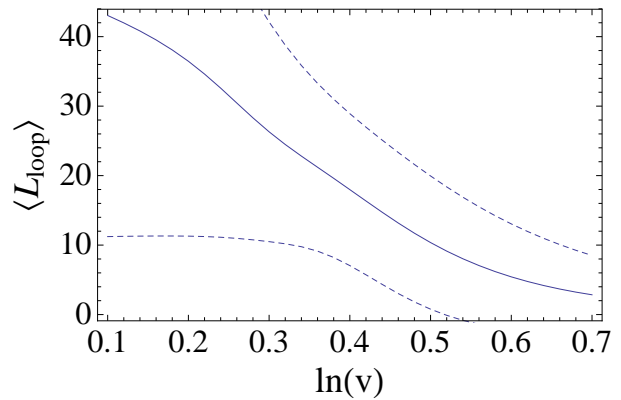}
\caption{The average loop length $\langle L_{loop}\rangle$ (solid line) is plotted
as a function of the per site binding strength of the chains using a bare
persistence length of the noninteracting chains of
$\ell_p=10a$.  The dotted lines provide a measure of the width of the
distribution of loop sizes by plotting
$\langle L_{loop}\rangle \pm \sqrt{\langle L_{loop}^2\rangle-\langle L_{loop}\rangle^2}$.}
\label{fig:loopfig}
\end{figure}

We consider a pair of bound polymers of length N in the limit
$N\rightarrow \infty$ so that end effects may be ignored. The
partition function for this system is
\begin{equation}
\lim_{N\to\infty}x_1^{2N}=\prod_{p=1}^\infty (U(1)V(1))^p.
\end{equation}
By translational symmetry all the loops are identical so we
calculate the $m$th moment of the length of the first loop from
\begin{eqnarray}
\langle L_{loop}^m\rangle&=&\frac{V(1)\left(\sum_{n=1}^\infty n^m
u(n)\right)\left(\prod_{p=0}^\infty (U(1)V(1))^p \right)}{x_1^{2N}}
\\
&=&V(x_1)\left(\sum_{n=1}^\infty n^m
\frac{u(n)}{x_1}\right)\left(\prod_{p=0}^\infty (U(x_1)V(x_1))^p
\right).
\end{eqnarray}
By the definition of $x_1$, $U(x_1)V(x_1)=1$, so we have
\begin{equation}
\langle L_{loop}^m\rangle=V(x_1)\left(\sum_{n=1}^\infty n^m
\frac{u(n)}{x_1}\right). \label{eq:loopaverage}
\end{equation}
In Fig \ref{fig:loopfig} we plot the average loop size along with its variance. The
loops are typically just a few persistence lengths as this length is
required to avoid a prohibitive bending energy penalty.  The narrow
distribution is indicates that longer length loops also incur a
prohibitive binding energy penalty.

We now turn to calculation of the effective persistence length of
the paired chains. The local bending modulus of the polymer changes
markedly when to chains are tightly bound. One should thus be able
to distinguish free loops from ``zipped'' regions of paired chains
purely from their local conformational statistics in equilibrium. We
first demonstrate this difference in persistence length by
estimating its value for free chains and tightly bound ones. In
their free state the persistence length of the polymers may be
inferred from the conjugation length observed in polymer melts.
Given a finite conjugation length of this form, one may simply
describe the conformational statistics of the free polymer as a
freely rotating chain having a segment length equal to its
conjugation length. This length $\ell_{min}$ is on the order of ten
monomers at for the case of PPV~\cite{Heun:93}, but
this value depends on the specific chemical system.

When the chains are tightly bound together in what we call the
``zipper'' state, the intermolecular bonds will prevent rotations
about the polymeric backbone. We imagine the principal remaining
source of backbone flexibility on the microscale to come from the
bending and stretching of the backbone chemical bonds. The energy
scales associated with bond deformation are significantly higher
than for the bond rotation of the free chains. The bond stretching
modes of the backbone are, in fact, too stiff to play a significant
role.  Using the estimation of the effective bond-stretching modulus
of 21 eV/\AA$^2$ we estimate that the effective paired-chain modulus
for bending perpendicular to the conjugation plane to be $\sim 120\
{\rm eV/radian}^2$ \cite{Su:80}. This is to be compared with the
$5.5\ {\rm eV/radian}^2$ modulus for bending the bonds in the
conjugation plane~\cite{Canales:03}. We estimate that effective
persistence length of the paired chain by determining length over
which a one radian uniform bend generates one $k_BT$ of elastic
energy and find that the paired-chain persistence length is at least
one order of magnitude bigger than that of the free chain, i.e.
$\ell_{max} \simeq 10 \ell_{min}$.

A representative equilibrium configuration of the paired chains
will, in general, consist of a series of alternating ``zipped''
regions and loops in which the two chains are locally unbound. Since
the ``zipped'' regions are significantly stiffer than the floppy
loops,  we may approximate the resulting configuration of the paired
chains as a random walk consisting of essentially straight
``zipped'' regions connected by flexible loop regions at which the
polymer bends freely.  To determine the effective persistence length
of the paired chains then we must determine the mean length of the
stiffer ``zipped'' regions in thermal equilibrium.  This length is
given by the ratio of fraction of bound sites $\theta$ to the
fraction of sites that make up the boundary of a ``zipped'' region:
$\theta_{\rm end}$. The latter value is calculated by appending a
boundary term to the loop regions such that $U(x)\rightarrow
\sigma_U U(x)$ that counts such boundaries. The fraction of sites
that incur this boundary weight is then given by
\begin{equation}
\theta_{\rm end}= \left. \frac{\partial \ln x_1}{\partial \ln
\sigma_U} \right|_{\sigma_U =1},
\end{equation}
which may be evaluated by implicitly differentiating
Eq.~\ref{xequation}. We obtain
\begin{equation}
\theta_{\rm
end}=\frac{\sigma_U}{x_1}\frac{-1}{\sigma_U\left(U\frac{\partial
V}{\partial x_1}+V \frac{\partial U}{\partial x_1}\right)}.
\end{equation}
The parameter $\theta$, determining the number of bound sites is
given by the derivative
\begin{equation}
\theta=\frac{\partial \ln x_1}{\partial \ln v}.
\end{equation}
Alternatively, $\theta$ could be calculated from an
expression analogous to Eq. \ref{eq:loopaverage}.
The average length of the ``zipped'' regions is
\begin{eqnarray}
\langle L_{\rm zip}\rangle&=&\frac{\theta}{\theta_{\rm end}}=
\frac{v}{V(x_1)}\frac{\partial V(x_1)}{\partial v} \nonumber \\
&=&\frac{vx_1^4+(v^3-2v^2x_1^2)(1-\sigma)}{(x_1^2-v)[vx_1^2-v^2(1-\sigma)]}.
\end{eqnarray}

The various regimes of the effective persistence length can be seen
in Fig.~\ref{fig:concfig} as a fraction of the persistence lengths
$\ell_{min},\ell_{max}$ and strength of the binding interaction as
parameterized by $\ln v$ -- see Eq.~\ref{v-definition}. Based on our
calculations, we expect this last parameter to be of order one.  In
this figure we see that the effective persistence length smoothly
crosses over from $\ell_{min}$ to $\ell_{max}$ as the binding
interaction increases for sufficiently long polymers, i.e. for those
polymers whose contour length $L$ is greater than $\ell_{max}$.

\section{Discussion}
\label{sec:conclusions}

We have further explored the effect of inter-chain electronic
tunneling on polymer pairing (i.e. forming two chain bound states)
and on the structure of those pairs in thermal equilibrium. At the
level of the formation of an isolated binding site, the binding
energy is highly insensitive to the density of states at the Fermi
energy of the respective chains. In particular, doped metallic
chains, explored earlier~\cite{Schmit:05}, and semiconducting ones
form individual binding sites that decrease the total electronic
energy of the system by $O(t'^2/t_0)\sim k_BT$. We also investigated
the electronic interaction between tunneling sites that are
separated by a short arc length along the chain. Here we see that
there is a weak attractive interaction between binding sites due to
the interaction of the localized states at these tunneling sites.
The strength of this attraction is, for at least one physical set of
model parameters~\cite{Bredas:02}, rather weak, $\sim 0.1 k_{B}T$,
but this value may vary widely between conjugated polymer systems.
The attractive interaction may in some cases lead to the clustering
of bound regions.

We support the single tunneling site results of our simple
tight-binding model using Hartree-Fock numerical calculations; the
numerics also show that the inter-chain binding is somewhat
insensitive to the precise orientation of the chains at the
tunneling site, although the angle between the chains at the
crossing point does lead to significant changes in the binding
energy. There we find that the perpendicularly crossed-state which
simultaneously allows for the a small separation at the tunneling
site and minimizes inter-chain steric repulsion is actually not the
lowest energy state. This configuration is a local minimum, while
the global minimum occurs for more nearly parallel chains in an
orientation consistent with the alternating binding and non-binding
sites corresponding to the ground state of the tight-binding model.
In the context of that model, we understand the formation of the
``every-other'' structured ground state in terms of enhancing the
Peierls splitting of the conduction and valence bands. It appears
from the Hartree-Fock calculations that the local structure of a
single binding site also favors the formation of such a
``every-other'' bound state. It is interesting to note that the
crystal structure of polyacetylene shows evidence of this
``every-other'' bonding pattern \cite{Winokur:87,Kahlert:87}. In this case, however,
each polymer in the crystal makes contact with several others, but
the local equivalence of the pattern of contacts suggests that the
electronic mechanisms we consider contribute to the stabilization of
this crystal structure.

The binding between two chains is enhanced by the co-localization of
solitons with the tunneling sites. The mixing of the mid-gap states
associated with the solitons by the tunneling matrix element leads
to this binding enhancement for uncharged solitons, but not for
charged ones. This effect is weak when there is only one soliton at
a tunneling site. Thus, solitonic binding enhancement should scale
as the square of the equilibrium (uncharged) soliton density.
Recalling that their density in thermal equilibrium is low, we do
not expect this to play a large role. It is important to note,
however, that a binding event between a charged soliton and an
uncharged soliton will result in one of the gap states being half
filled and therefore lead to half of the total possible binding
enhancement. The density of charged solitons is easily controlled by
doping. Using this, one could, in the presence of a sufficient
density of charged solitons, produce a significant solitonic
enhancement (on the order of $1 k_B T$ per site) of the chain
binding energy that scales linearly with the density of uncharged
solitons. We expect the same behavior to be observable in
polaron/tunneling site interactions in conjugated polymers more
chemically complex than polyacetylene.

The statistical mechanics of two paired conjugated polymers is
rather subtle since the binding sites generate free energy changes
of only $\sim 1 k_B T$ per site and allow for significant local
conformational freedom at a binding site. The structure of the
paired chains results from a nearly equal competition of chain
configuration entropy and binding energy. Of course, many of the
same subtleties have been addressed in the problem of DNA melting.
The one main difference between DNA melting and the binding of
conjugated polymers, is that in the latter there is no equivalent of
the base-pairing mechanism that promotes the binding in registry of
complementary strands. The lack of pairing registry in the current
problem enhances the entropic gain of loop-formation by allowing
loops of a fixed length $L$ to be created from various lengths $l_1$
and $l_2$ (such that $l_1 + l_2 = L$) from the two chains. Using a
modified Poland-Scheraga model which accounts for this difference,
we found that the effective persistence length varies over at least
an order of magnitude as a result of inter-chain binding. The bound
regions are significantly stiffer than the unpaired or free loops of
the chains allowing us to calculate the effective persistence length
of the paired chain in terms of the mean length of the stiff, bound
or ``zipped'' regions.

A number of open questions remain. The spectroscopic signature of
the pairing mechanism on two chains is still unresolved. Moreover,
the formation of polymeric aggregates of more than two chains
remains an open question, particularly with regard to the formation
of crystals of conjugated polymers. The fundamental tight-binding
approach to the electronic degrees of freedom coupled to the
statistical mechanics of the chains, further developed in this
article, should provide the basis for these future investigations.

\acknowledgements The authors would like to thank F.\ Pincus, G.\
Bazan, and A.J.\ Heeger for stimulating conversations. JDS would
also like to thank D.\ Scalapino and L.\ Balents for helpful
discussions. JDS acknowledges the hospitality of the University of
Massachusetts, Amherst.  This work was supported in part by the
MRSEC Program of NSF DMR00-80034 and NSF DMR02-03755.

\appendix

\section{Transfer Matrix Calculations}
\label{apdx:matrix}

The one dimensional character of the electronic hopping mechanism
lends itself naturally to a transfer matrix approach. In this
appendix we review the formalism as it is used to solve for the
electronic eigenstates of interacting polymers.  In the unperturbed
state the eigenstates are normal Bloch waves. The introduction of
the inter-chain interaction will scatter these waves and lead to
localized states. We begin from the electronic part of the
interacting Hamiltonian, Eqs.~\ref{sshH} and \ref{interaction}, and
work in the basis of chain symmetrized states as shown in
Eq.~\ref{symmetrize}. After this transformation, the symmetric and
anti-symmetric parts of the Hamiltonian may be written
\begin{eqnarray}
H_{S/A}&=&-\sum_{\ell,\sigma}t_{\ell,\ell+1}(\ket{\ell+1,\sigma}
\bra{\ell,\sigma}+\ket{\ell,\sigma}\bra{\ell+1,\sigma}) \nonumber \\
&&+\lambda \ket{0,\sigma}\bra{0,\sigma}, \label{eq:onesite}
\end{eqnarray}
where $\lambda=-t'(+t')$ in the symmetric (anti-symmetric)
sub-space.

We solve for the energy eigenvalues $E$ and the eigenvectors
$\ket{E}=\sum_\ell c_\ell \ket{\ell}$ that satisfy the Schrodinger
equation $H_{S/A}\ket{E}=E\ket{E}$.  By operating on the Schrodinger
equation with the vector $\bra{\ell}$ we find that
the amplitudes $c_\ell$ obey the set of simultaneous equations
\begin{equation}
-t_{\ell-1,\ell}c_{\ell-1} + \varepsilon_\ell c_\ell - t_{\ell,\ell
+ 1}c_{\ell+1}=E c_\ell,
\end{equation}
which, along with a trivial identity, may be written in matrix form.
\begin{equation}
\left(\begin{array}{cc}
0 & 1 \\
-\frac{t_{\ell-1,\ell}}{t_{\ell,\ell+1}} & \frac{\varepsilon_\ell
-E}{t_{\ell,\ell+1}}
\end{array}\right)\left(
\begin{array}{c}
c_{\ell-1} \\ c_\ell \end{array}\right)=\left(\begin{array}{c}
c_\ell \\ c_{\ell+1} \end{array}\right). \label{eq:transfer}
\end{equation}
From Eq.~\ref{eq:onesite} the on-site energies $\varepsilon_\ell$
vanish everywhere except at the binding site,
$\varepsilon_\ell=\delta_{0,\ell}\lambda$.

To bring the Hamiltonian into Bloch form, we construct the transfer
matrix for a unit cell
\begin{equation}
M_{\ell+1}M_{\ell}= \left(\begin{array}{cc}
-\frac{t_2}{t_1} & -\frac{E}{t_1} \\
\frac{E}{t_1} & -\frac{t_1}{t_2}+\frac{E^2}{t_1 t_2}
\end{array}\right).
\label{doubleM}
\end{equation}
If we number the carbons according to 1=2--3=4--5\ldots, where the
single and double lines represent long and short bonds respectively,
Eq.~\ref{doubleM} propagates of the wavefunction between consecutive
odd sites.  As required by the discrete translational invariance of
the Hamiltonian, Eq.~\ref{doubleM} takes a diagonal form in the
momentum basis
\begin{equation}
UM_{\ell+1}M_{\ell} U^{-1}= \left(\begin{array}{cc} e^{i2ka} & 0
\\ 0 & e^{-i2ka} \end{array}\right),
\label{diagonalize}
\end{equation}
where $2a$ is the unit cell dimension of the
dimerized chain, and
\begin{eqnarray}
e^{\pm i2ka}&=&\frac{E^2-t_1^2-t_2^2}{2t_1 t_2}  \nonumber \\
&&\pm \sqrt{\left(\frac{E^2-t_1^2-t_2^2}{2t_1 t_2}\right)^2-1} \label{kdef}\\
U&=& \frac{1}{2i\sin(q)}\left(\begin{array}{cc} 1 & -e^{-iq} \\ -1 &
e^{iq} \end{array}\right).
\end{eqnarray}
The first of these equations implicitly redetermines the dispersion
relation of the Bloch waves making up the energy eigenstates of the
system. The solution $E(k)$ is given below in
Eq.~\ref{apdx-dispersion}. For completeness we also report the
change of basis matrix defined by Eq.~\ref{diagonalize} in terms of
the dimensionless parameter $q$ given by
\begin{equation}
e^{iq}= \frac{-E}{t_1 e^{i2ka}+t_2} \label{qdef}.
\end{equation}
The complex number is the phase change of the wavefunction
associated with the quantum number $k$ (wavenumber) across the long
bonds separating one unit cell from the next. In other words:
$c_\ell=c_{\ell+1}e^{iq}$, as may be checked using
Eqs.~\ref{eq:transfer}, \ref{doubleM}, and \ref{qdef}.  The phase
shift within a unit cell is $2ka-q$ so that the total phase shift
from one unit cell to the next is $2ka$, as required by Bloch's
theorem.

Of course, the condition for bounded eigenfunctions is that $k$ must
be real, leading to the dispersion relation
\begin{equation}
\label{apdx-dispersion} E(k)=\pm\sqrt{t_1^2+t_2^2+2t_1 t_2
\cos(2ka)}
\end{equation}
which has a total bandwidth of $4t_0$ and a bandgap centered around
$E=0$ of width $8u_0\alpha\equiv2\Delta$.

To study the effect of the binding site on the electronic spectrum
we repeat the transformation Eq. \ref{diagonalize} with the dimer
containing the binding site at $\ell=0$.  The result is
\begin{equation}
\label{Mscatter} M_\lambda=UM_1M_{0} U^{-1}=\left(\begin{array}{cc}
e^{i2ka}(1+i\delta) & i\delta e^{i2ka} \\ -i\delta e^{-i2ka} &
e^{-i2ka}(1-i\delta)\end{array}\right),
\end{equation}
where $\delta=\lambda E/2t_1 t_2 \sin(2ka)$.  In addition to the
states given by Eq.~\ref{eq:dispersion} for $-\pi<2ka<\pi$, the
off-diagonal elements in Eq.~\ref{Mscatter} allow normalizable
states with complex wavenumber through the scattering of a growing
wave into a decaying one. This is accomplished when the upper left
matrix element vanishes, therefore
\begin{eqnarray}
1&=&-i\delta \\
&=& \frac{-i\lambda E_b}{2 t_1 t_2 \sin(2ka)} \\
&=&\frac{\lambda E_b}{\sqrt{(E_b^2-t_1^2-t_2^2)^2-4 t_1^2t_2^2}}.
\end{eqnarray}
With a little more algebra we find that the four bound states shown
in Fig.~\ref{fig:crossmodel}c are given by
\begin{equation}
E_b=\pm{\rm
sign}(\lambda)\sqrt{t_1^2+t_2^2+\frac{\lambda^2}{2}\pm\sqrt{4t_1^2
t_2^2+\lambda^2(t_1^2 +t_2^2)+\frac{\lambda^4}{4}}},
\end{equation}
where the upper sign applies to the ``ultraband'' states
(Eq.~\ref{eq:boundstatesU}) and the lower sign applies to the gap
states (Eq.~\ref{eq:boundstatesG})~\cite{Phillpot:87}.

We now compute the analogous bound state associated with a tunneling
site localized at the center of an idealized soliton.  We represent
the soliton as a single-site disruption of the dimerization pattern
as discussed in Section~\ref{sec:solitons}. It may be written as:
\linebreak $\circ=\circ-\circ=\circ-\bullet-\circ=\circ-\circ=\circ$
and has undistorted unit cells on either side of the soliton's
center ($\bullet$) connected to that special site by the long bonds
($-$) that have tunneling matrix element $t_2$. The tunnel matrix
element also occurs at this special and central site.

The transfer matrix at the central site is
\begin{equation}
M_s=\left(\begin{array}{cc}
0 & 1 \\
-1 & \frac{\lambda -E}{t_2}
\end{array}\right).
\end{equation}
Upon changing basis as shown in Eq.~\ref{diagonalize} we find
\begin{eqnarray}
&&UM_s U^{-1}=\frac{1}{2i\sin(k)}\times \\
&&\left(
    \begin{array}{cc}
    2-e^{-iq}\frac{\lambda -E}{t_2} & 1+e^{-2iq}-e^{-iq}\frac{\lambda -E}{t_2} \\
    -1-e^{2iq}+e^{iq}\frac{\lambda -E}{t_2} & -2+e^{iq}\frac{\lambda -E}{t_2}
    \end{array}\right). \nonumber
\end{eqnarray}
Once again, the bound states exist for values of $E$ where the upper
left element vanishes, so we have
\begin{eqnarray}
1=e^{-iq}\frac{\lambda-E_b}{2t_2}. \label{eq:solitonroot}
\end{eqnarray}
Using Eqs.~\ref{kdef},~\ref{qdef} we find
\begin{equation}
1=\frac{E_b^2-\lambda
E_b}{E_b^2-t_1^2+t_2^2-\sqrt{(E_b^2-t_1^2-t_2^2)^2-4t_1^2t_2^2}}.
\end{equation}
The solutions are given by the roots of the polynomial
\begin{equation}
0=E_b^4-E_b^2(4t_0^2+\Delta^2+\lambda^2)+4E_b\lambda t_0\Delta.
\end{equation}
It is easily verified that the solution $E_b=0$ does not satisfy
Eq.~\ref{eq:solitonroot}, and thus we arrive at
Eq.~\ref{eq:solitondispersion}.

\section{Hartree-Fock Calculations}
\label{sec:hartree}

To better understand the limitations of the tight-binding model and
explore the role of steric interactions near the tunneling site for
the specific case of polyacetylene we numerically computed the
binding energies of two short (20 monomers each) polyacetylene
oligomers with quantum chemical calculations using the package
Gaussian~\cite{Gaussian}.  After relaxing a single chain of fixed
length, we introduced an identical chain in various geometries. In
each case the interaction energy was calculated as the energy of the
two chain system minus twice the energy of the single chain.
Energies were evaluated using the non-counterpoised Hartree-Fock
technique with the 6-31G* basis set.  Additional calculations using
the B3LYP method found results qualitatively similar to those from
Hartree-Fock (data not shown).
\begin{figure}[ht]
\vspace{0.6 cm}
\begin{center}
\includegraphics[width=6.0cm]{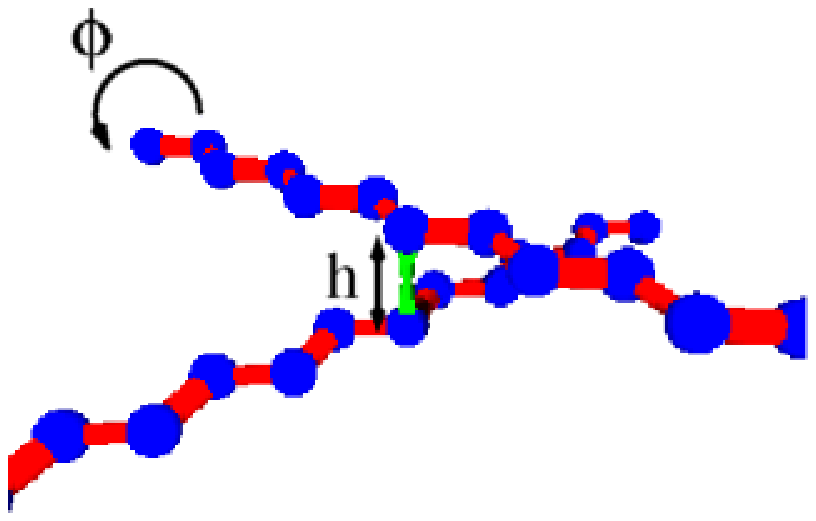}
\includegraphics[width=6.0cm]{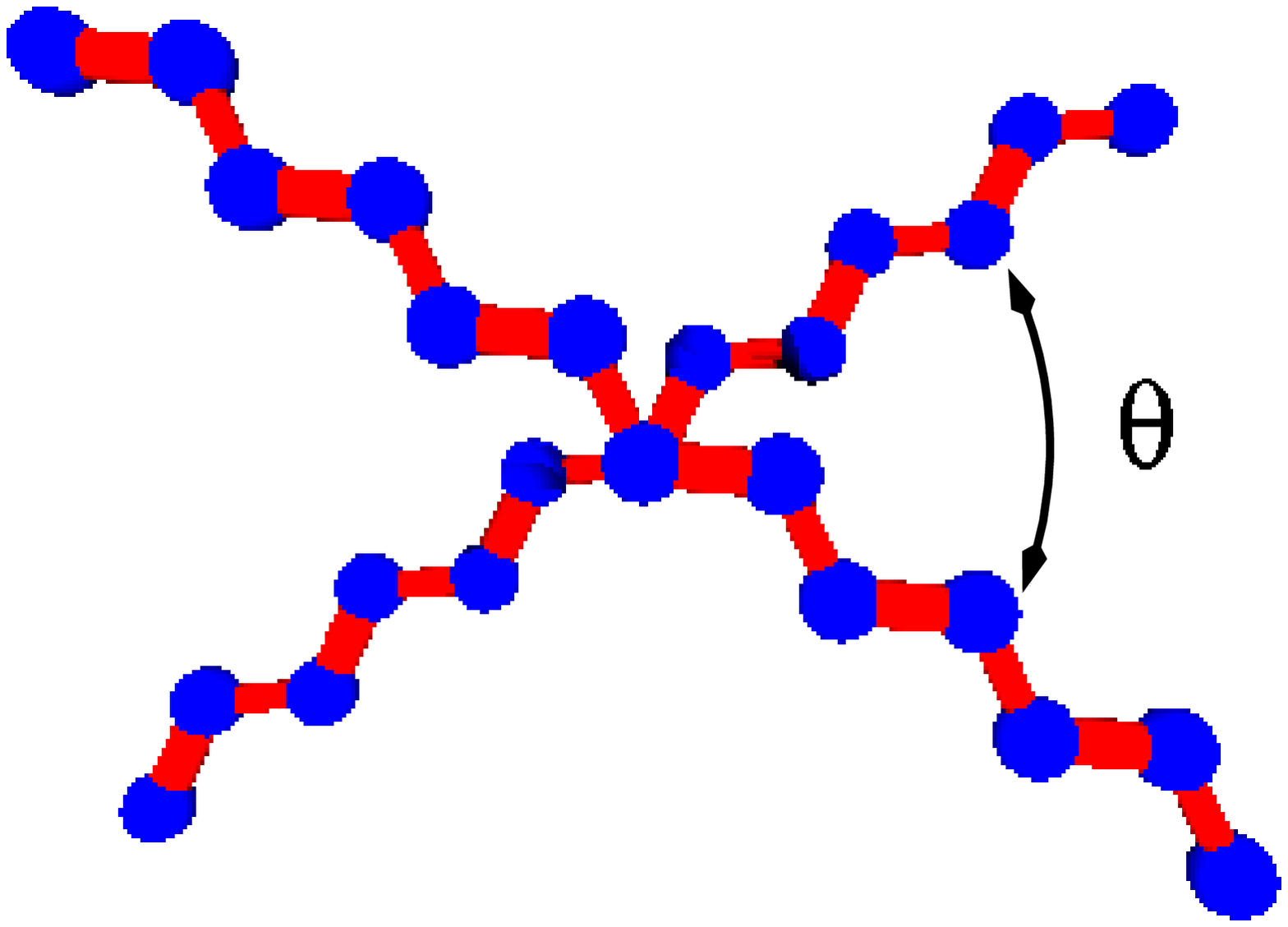}
\end{center}
\caption{ \label{fig:diagrams} (top) Two polyacetylene chains from
the view (top) showing the vertical separation between the chains
$h$ and axial rotation angle $\phi$. The tunneling site is shown as
the (green) pillar. (bottom) The same chains from the top view,
looking down along the direction of $h$ and demonstrating the
crossing angle $\theta$.}
\end{figure}

\begin{figure}[ht]
\vspace{0.6 cm}
\begin{center}
\includegraphics[angle=0,scale=0.75]{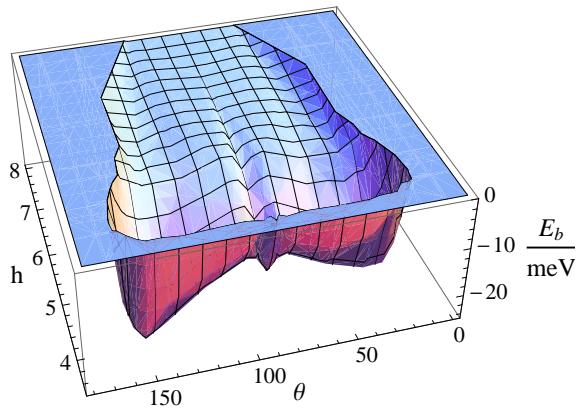}
\end{center}
\caption{ Binding energy between two 20 carbon polyacetylene chains
as a function of their separation $h$ and contact angle $\theta$.
The optimal separation is $h=4.0$\AA\ at all angles. There are
pronounced local minima at $\theta=30^\circ$ and $\theta=90^\circ$
and a global minimum at $\theta=150^\circ$.  The flat surface
represents regions where the net binding energy is repulsive within
the HF approximation. \label{fig:hbind}}
\end{figure}

Using this procedure we tested the distance and orientation
dependence of the binding energy due to a single tunneling site. We
parameterized the geometry of the near miss of two chains by two
angles and one distance. As shown in Fig.~\ref{fig:diagrams}, the
perpendicular distance between the two chains at the binding site is
$h$. The angle $\phi$ measures the rotation of the plane containing
the intra-chain carbon bonds of one polymer about an axes parallel
to that chain. Finally, the angle between the long axes of the two
chains (looking down along the direction parallel to the vertical
displacement $h$) is given by $\theta$. In Fig.~\ref{fig:hbind} we
show the binding energy obtained numerically as a function of $h$
and $\theta$ for a fixed value of $\phi = 0$ (i.e. with the
conjugation planes parallel). As expected for the tunneling
mechanism, the attractive interaction decays rapidly (exponentially)
with distance reaching minimum at $h=4.0$ \AA\ for all values of
$\theta$.
The angular dependence at fixed $h$ is more complex. For $h=4.0$\AA\
corresponding to maximum binding, the attractive interaction has a
minimum at the perpendicular crossing of the chains ($\theta=\pi/2$)
where the net binding energy is $-0.0137$ eV. This local minimum is
globally unstable; there are deeper minima near $\theta = \pi/6,
5\pi/6$ corresponding to more nearly parallel alignments of the
chains. The $\theta=5\pi/6$ minimum is deepest and in closer
agreement to the ``every other'' site ground state predicted by the
tight binding model. For precisely parallel alignments ($\theta=0$)
in which the chains are stacked on top of each other as shown in
Fig.~\ref{fig:fullzipfig}, they do not attract, as predicted by our
tight binding model.  In fact, interactions not included in our
model give a repulsive interaction of $0.154$ eV. Our results are
consistent with previous work that has shown the maximally
overlapped configuration to be a high energy state with more
favorable interactions occurring when one of the molecules is
rotated or translated such that the overlap is removed from some of
the carbons~\cite{Sinnokrot:06,Hutchinson:05}.

The binding energy shown in Fig.~\ref{fig:hbind} is also insensitive
to the precise lateral alignment of the two chains at the binding
site.  We shifted the top chain by $0.2$ \AA\ in the positive and
negative X and Y directions, where $h$ lies along the Z direction.
These shifts altered the binding energy by less than 6\%. These data
are not shown.

\begin{figure}[ht]
\vspace{0.6 cm}
\begin{center}
\includegraphics[angle=0,scale=1.0]{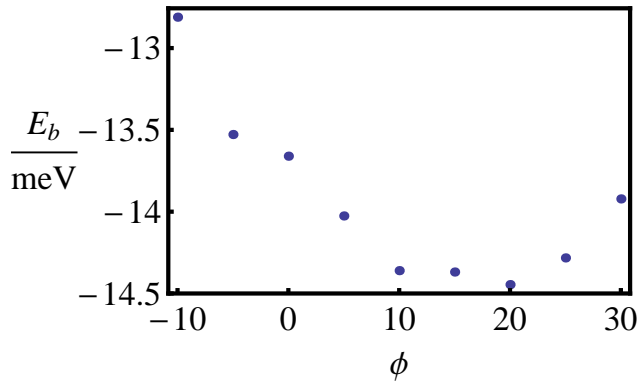}
\end{center}
\caption{ \label{fig:phibind} The interaction energy as a function
of $\phi$ for $h=4.0$ \AA\ and $\theta=90^\circ$.}
\end{figure}

Finally, we rotated the upper chain by the angle $\phi$ as defined
in Fig.~\ref{fig:diagrams}.  This rotation preserved the $4$ \AA\
separation between the two carbons at the binding site. For negative
rotations the interaction becomes less favorable due to the steric
repulsion of the nearby carbons of the two chains, but for positive
rotations the binding energy increases to $0.0144$ eV at an angle of
$\phi = 0.3$, as shown in Fig.~\ref{fig:phibind}. Note that these
energy shifts are significantly less than $k_{B}T$ suggesting that
the rotational motion of the polymers is unconstrained by the
tunneling sites in thermal equilibrium.

We find that these results are only weakly dependent upon chain
length; for example, as the chain lengths were increased from ten to
forty carbons, the depth of the well increased by about five
percent. In order for the numerically obtained binding energies to
match our tight binding results, we need to set $t_0\simeq 2.5$ eV
and $t'\simeq 0.26$ eV. This result is consistent with the overlap
measured in Ref.~\cite{Bredas:02} for configurations where the
contact extends over the entire molecule.

The Hartree-Fock calculations presented here support the notion of
intermolecular bonds forming at the close approach of two
polyacetylene chains. The strength of these interactions is
consistent with previous studies~\cite{Bredas:02}. The range of the
attraction is as expected for the proposed tunneling mechanism and
its angular ($\theta$) dependence suggests both perpendicular and
nearly parallel arrangements are favored for a single binding site.
The binding interaction, however, is sufficiently insensitive to the
precise orientation of the polymers as to allow the polymers to form
a variety of aggregates.

We conclude by noting that the {\it ab initio} calculation of the
interactions between $\pi$-conjugated molecules is an area of
ongoing research \cite{Sinnokrot:06,Sherrill:09}.  While significant
advances have been made for small aromatic molecules, these
molecules lack the extended orbitals essential to our model.  These
methods are not feasible for the larger systems we require, and
therefore, we have employed the significantly less accurate HF
method.  While we are encouraged by the qualitative agreement with
our tight-binding model, we emphasize that our HF results should not
be viewed as quantitative.

\section{Binding Site Interactions}
\label{apdx:siteinteract}

To examine the interaction energy between two tunneling sites we
consider a chain of $N$ tight binding sites.  We place the two
tunneling sites, which act like impurity potentials, at sites $M$
and $M+d$ such that the the tunneling sites are symmetrically placed
about the center of the chain: $N=2M+d-1$. This provides a
computational convenience without greatly reducing the validity of
the results so obtained. The position of the \emph{localized states}
along the chain should be irrelevant on long polymers $N \gg 1$
provided neither tunneling site is within its localization length (a
few lattice constants) of the chain ends.

\begin{figure}[htpb]
\centering
\includegraphics[width=8.0cm]{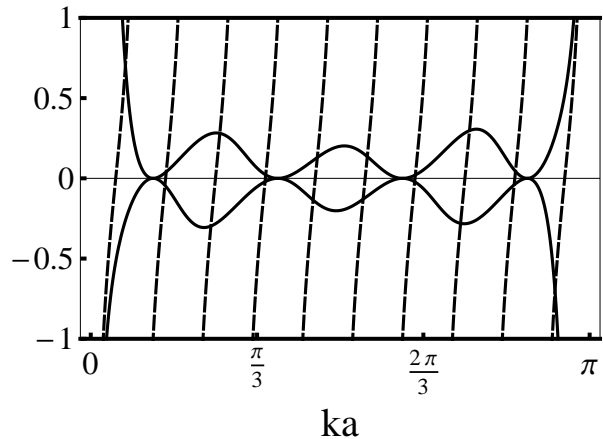}
\caption{Graphical solution of Eq. \ref{evenquant} (see text).}
\label{fig:twositequantplot}
\end{figure}

Since the even-odd effect observed in Fig.~\ref{fig:twositesD} is
unchanged by the chain dimerization, we consider the case of a
uniform chain with $t_{\ell,\ell+1}=t_0$ and dispersion
$E(k)=-2t_0\cos (k)$.  The eigenstates have the form
\begin{eqnarray}
\ket{k}&=&\sum_{\ell=1}^{M-1} \sin (k \ell) \ket{\ell} \nonumber \\
&&+A\sum_{\ell=L}^{M+d} \sin (k \ell+\phi) \ket{\ell} \nonumber \\
&&\pm\sum_{\ell=M+d+1}^{N} \sin (k (N+1-\ell)) \ket{\ell}
\end{eqnarray}
where $A$ is an amplitude to be determined, and the sign of the
final term is determined by parity.  The boundary conditions require
that the wavefunction be an eigenstate of the parity operator and
and remain continuous at the impurities so that
\begin{eqnarray}
\label{twositeparity}
&kM+\frac{kd}{2}+\phi=\frac{n\pi}{2}& \\
\label{twositecont}
&\sin (kM)=A \sin (kM+\phi).& \\
\end{eqnarray}
We derive a final condition by requiring that the Schr\"{o}dinger
equation is satisfied at the impurity sites
\begin{eqnarray}
&&-2t_0 \cos(k) \sin(kM)=\lambda\sin(kM) \nonumber \\
 &&-t_0\sin[k(M-1)] -t_0 A\sin[k(M+1)+\phi].
\label{twositeSE}
\end{eqnarray}
Using Eqs. \ref{twositeparity}, \ref{twositecont}, and
\ref{twositeSE} we find the conditions on $k$ for states of even and
odd parity are
\begin{eqnarray}
\frac{\lambda}{t_0 \sin(k)}=\tan (\frac{kd}{2})-\cot(kM)&&{\rm even\ states} \\
\frac{\lambda}{t_0 \sin(k)}=-\cot (\frac{kd}{2})-\cot(kM)&&{\rm odd\ states}.
\end{eqnarray}
We would like to examine the change in energy of the states as the
distance between the impurity potential is changed but the length of
the chain is constant.  To do this we make the substitution
$M=(N-d+1)/2$ and find that the quantization conditions become
\begin{eqnarray}
\label{evenquant}
-\cot \left( k
\frac{N+1}{2}\right)=\frac{\frac{\lambda}{2t_0}[1+\cos(kd)]}{\sin(k)-
\frac{\lambda}{2t_0}\sin(kd)}&&{\rm
even} \\
\tan \left( k
\frac{N+1}{2}\right)=\frac{\frac{\lambda}{2t_0}
[1-\cos(kd)]}{\sin(k)+\frac{\lambda}{2t_0}\sin(kd)}&&{\rm
odd}.
\end{eqnarray}
To see the energy shift of the individual states it is helpful to
graphically solve these equations.  In
Fig.~\ref{fig:twositequantplot} we plot both sides of
Eq.~\ref{evenquant} for the symmetric and antisymmetric states
($\lambda/2t_0=\pm 0.5$) and $d=8$. The left side of the equation,
drawn in dashed lines, is a series of tangent curves with a spacing
set by the length of the chain.  For very long chains these tangent
curves will be spaced infinitesimally close together. The allowed
$k$ values are found at the intersections of the solid curves and
the dashed curves. In the limit $\lambda \rightarrow 0$, the solid
curves lie along the $k$ axis and we find the unperturbed states
satisfy $k=(2n+1) \pi /2(N+1)$ where $n$ is an integer.  For small
positive values of $\lambda$ the right side of the equation takes
positive values and the eigenstate is shifted to larger $k$ values.
Similarly, negative values of $\lambda$ shift the states to longer
wavelengths.  Therefore, the amplitude of the solid curves in
Fig.~\ref{fig:twositequantplot} grows with the magnitude of the
wavenumber shift of the eigenstates.

We see in Fig.~\ref{fig:twositequantplot} that, although the curves
for positive and negative values of $\lambda$ oscillate with the
same period, the curves are not symmetric about the $k$ axis.
Starting at small $k$, there are alternating regions where the
symmetric states are shifted more than the anti-symmetric states and
vice versa.  The origin of this asymmetry between the sub-spaces is
the phase shift induced by the impurity potentials discussed in
section~\ref{twositesection}. The contribution of these states to
the total energy of the system will depend on the location of the
Fermi level.  If the separation between the binding sites, $d/a$, is
even the Fermi level, $E_f=0$, lies just above a region of large
shift for the symmetric states and below the corresponding region of
large shifts for the anti-symmetric states.  This results in a net
negative contribution to the binding energy.  However, if the
separation is odd the Fermi level lies immediately above a region of
large shifts for the anti-symmetric states and the contribution is
positive.


\begin{thebibliography}{Bredas:02}
\bibitem{phaseshift} An electronic state scattering off an impurity potential, $\lambda$
at the origin has the form $\ket{k}=\sum_{l\leq 0} \cos (kla
+\theta)\ket{l}+A\sum_{l> 0} \cos (kla +\theta+\phi)\ket{l}$.  Using
the requirements that the wave satisfy the Schr\"{o}dinger equation
and is continuous at the impurity site, we solve for the
undetermined amplitude, $A$, and the phase shift
$\phi=\tan^{-1}\left(\tan(\theta)-\frac{\lambda}{t\sin(ka)}\right)-\theta$.
\end{thebibliography}
\end{document}